\documentclass[iop]{emulateapj}

\slugcomment{Accepted to the Astrophysical Journal: August 10, 2014 issue}

\usepackage{graphicx}
\usepackage{natbib}

\shorttitle{Abundant molecular gas and inefficient star formation in intracluster regions}
\shortauthors{J\'achym et al.}

\begin{document}
\title{Abundant molecular gas and inefficient star formation in 
intracluster regions: Ram pressure stripped tail of the Norma galaxy 
ESO137-001\altaffilmark{*}}

\author{Pavel J\'achym\altaffilmark{1}, 
Fran\c coise Combes\altaffilmark{2}, 
Luca Cortese\altaffilmark{3}, 
Ming Sun\altaffilmark{4}, 
Jeffrey D. P. Kenney\altaffilmark{5}
}
\altaffiltext{1}{Astronomical Institute, Academy of Sciences of the Czech 
Republic, Bo\v cn\'i II 1401, 14100, Prague, Czech Republic, \email{jachym@ig.cas.cz}}
\altaffiltext{2}{Observatoire de Paris, LERMA, 61 Av. de l'Observatoire, 75014 
Paris, France}
\altaffiltext{3}{Centre for Astrophysics \& Supercomputing, Swinburne University 
of Technology, Mail H30, P.O. Box 218, Hawthorn, VIC 3122, Australia}
\altaffiltext{4}{Department of Physics, University of Alabama in Huntsville, 301 
Sparkman Drive, Huntsville, AL 35899, USA}
\altaffiltext{5}{Department of Astronomy, Yale University, 260 Whitney Ave., New 
Haven, CT 06511, USA}
\altaffiltext{*}{Based on observations made with ESO telescopes at La Silla 
Paranal Observatory under programme ID~088.B-0934.}

\begin{abstract}
For the first time, we reveal large amounts of cold molecular gas in a 
ram pressure stripped tail, out to a large, ``intracluster'' distance 
from the galaxy. With the APEX telescope we have detected 
$^{12}$CO(2-1) emission corresponding to more than $10^9~M_\odot$ of 
H$_2$ in three H$\alpha$ bright regions along the tail of the Norma 
cluster galaxy ESO~137-001, out to a projected distance of 40~kpc from 
the disk. ESO~137-001 has an 80~kpc long and bright X-ray tail 
associated with a shorter (40~kpc) and broader tail of numerous 
star-forming \ion{H}{2} regions. The amount of $\sim 1.5\times 
10^8~M_\odot$ of H$_2$ found in the most distant region is similar to 
molecular masses of tidal dwarf galaxies, though the standard 
Galactic CO-to-H$_2$ factor could overestimate the H$_2$ content.
Along the tail, we find the amount of molecular gas to drop, while 
masses of the X-ray emitting and diffuse ionized components stay 
roughly constant. Moreover, the amounts of hot and cold gas are large 
and similar, and together nearly account for the missing gas from the 
disk. We find a very low star formation efficiency ($\tau_{\rm dep}> 
10^{10}$~yr) in the stripped gas in ESO~137-001 and suggest that this 
is due to a low average gas density in the tail, or turbulent heating 
of the interstellar medium that is induced by a ram pressure shock. 
The unprecedented bulk of observed H$_2$ in the ESO~137-001 tail 
suggests that some stripped gas may survive ram pressure stripping in 
the molecular phase. 
\end{abstract}


\section{Introduction}
The dense environments of galaxy clusters and groups have been
identified as places where transformations of galaxies from blue,
star-forming to red, gas-poor systems happen. Late-type galaxies are
rare in the cores of galaxy clusters although dominate in their outer
parts and outside of clusters \citep[the morphology-density
relation,][]{dressler1980,vanderwel2010}. Cluster galaxy populations
evolve over cosmic time, as shown by the excess of optically blue
galaxies in clusters at higher redshifts \citep[the Butcher-Oemler
effect,][]{butcher1978, butcher1984}. Several mechanisms are active in
clusters that could account for the observed evolution, such as mutual
gravitational interactions of galaxies, including mergers and rapid galaxy
encounters, alias harassment, the tidal influence of the cluster
potential, and ram pressure of the intracluster medium (ICM) on the
interstellar matter (ISM) of member galaxies \citep{gunn1972},
accompanied by numerous hydrodynamic effects.

While there is a wealth of observational evidence of one-sided, clearly
ram pressure stripped, gas tails in:
(a) \ion{H}{1} in the Virgo cluster \citep[e.g.,][]{kenney2004,
chung2007, chung2009, abramson2011, kenney2014} and more distant
clusters \citep[A1367,][]{scott2010, scott2012},
(b) diffuse H$\alpha$ tails \citep{gavazzi2001, cortese2006,
cortese2007, sun2007, yagi2007, yoshida2004, yoshida2008, kenney2008,
fossati2012},
(c) young stars seen either in H$\alpha$ or UV
\citep{cortese2006, sun2007, yoshida2008, smith2010, hester2010,
yagi2013, ebeling2014}, and
(d) X-rays \citep{wang2004, finoguenov2004, machacek2005, sun2005, sun2006,
sun2010},
\textit{no example} of a prominent (cold) molecular one-sided tail is
known up to now, except a few cases of extraplanar molecular gas
located fairly close to the disk \citep[NGC~4438,
NGC~4522,][]{vollmer2005, vollmer2008}. 
Recently, \citet{jachym2013} searched for cold molecular gas in the 
star-forming ram pressure stripped tail of the Virgo dwarf galaxy 
IC3418, that is (almost) completely stripped, but reached only 
sensitive upper limits.
Presence of molecular gas has been known in some intracluster regions 
though, as revealed by Carbon monoxide (CO) emission, the most widely 
used tracer of the molecular interstellar medium. Observations of 
filaments near cluster cores have revealed anomalously strong H$_2$ 
emission that may extend to large distances from the cD galaxies 
\citep{johnstone2007, salome2011}. Also, in the off-disk locations of 
the Virgo cluster galaxy M86, \citet{dasyra2012} found CO emission in 
the long H$\alpha$ tidal trail that connects M86 with NGC~4438. 

One of the main issues in galaxy formation and evolution is to
understand the star formation history, and in particular how star
formation is triggered, and how it can be suddenly quenched.
Environmental effects could play an important role in this matter, and
in particular ram pressure stripping of the gas of spiral galaxies.
Stars form from cold molecular interstellar gas gathered into giant
molecular clouds \citep[GMCs, e.g.,][]{wong2002}.
While \ion{H}{1} can be easily stripped from galaxies by the ICM wind,
the GMCs are less efficiently removed since they are
typically denser and distributed with a higher concentration to the
galactic center. Very recently, \citet{boselli2014} found evidence that
Virgo cluster galaxies that are \ion{H}{1} deficient have on average
also somewhat less molecular gas than \ion{H}{1}-normal field galaxies. The
effective H$_2$ ram pressure stripping in these galaxies may be a
consequence of removal of diffuse \ion{H}{1} that normally feeds giant
molecular clouds. Moreover, hydrodynamic ablation by the ICM wind is
assumed to assist the effective H$_2$ stripping \citep{nulsen1982,
quilis2000, kapferer2009, tonnesen2009}.

Numerical simulations predict that in the gas-stripped tails, depending 
on the ICM ambient pressure, radiative cooling can form in situ new 
dense (molecular) clouds which might then form a population of
new stars \citep{kapferer2009, tonnesen2009}. If their native gas 
clumps were accelerated by ram pressure to velocities exceeding the 
escape speed from the galaxy, such stars can contribute to the 
intracluster light (ICL) population. In extreme environments of massive 
or merging galaxy clusters it is possible that strong ram pressures 
could directly strip the dense molecular clouds.

\subsection{The galaxy ESO~137-001}
\begin{figure}[t]
\centering
\includegraphics[width=0.46\textwidth]{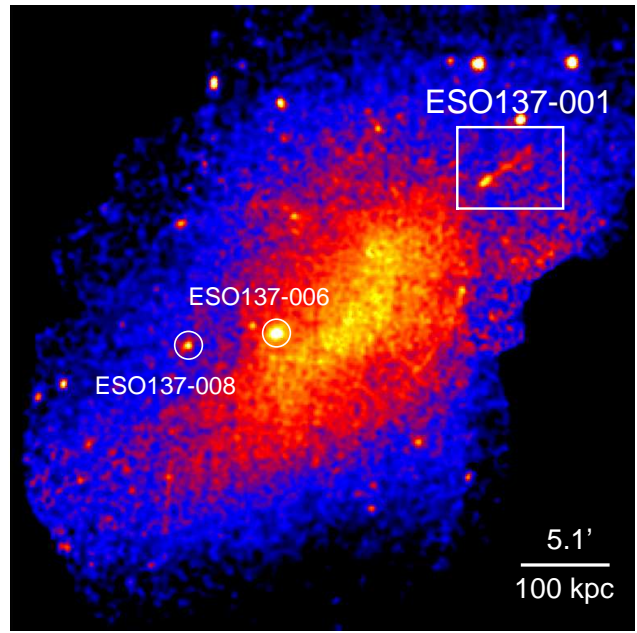}
\caption{
Central part of the A3627 cluster in a $0.5-2$~keV XMM image. The left
circle marks position of ESO~137-008 and the right one of ESO~137-006, 
two galaxies in the central region of the cluster with the same 2MASS
K$_{\rm s}$-band magnitude. Position of ESO~137-001 is marked with a 
rectangle. We zoom in into this region in Fig.~\ref{FigPoints}.
}\label{FigNorma}
\end{figure}

The Norma cluster (Abell 3627, $R_{\rm A}= 2$~Mpc, $M_{\rm dyn}\sim 
1\times 10^{15}~M_\odot$, $\sigma= 925$~km\,s$^{-1}$) is the closest
($z= 0.0163$) rich cluster comparable in mass and number of galaxies to
Coma or Perseus\footnote{Mean radial velocity of $4871\pm 
54$~km\,s$^{-1}$ was measured for the A3627 cluster by
\citet{woudt2008} from radial velocities of 296 member galaxies.}. It
occurs close to the center of the Great Attractor,
at the crossing of a web of filaments of galaxies \citep[dubbed the
Norma wall;][]{woudt2008}. The Norma cluster is strongly elongated along
the Norma wall indicating an ongoing merger at the cluster core (see
the central part of the cluster in Fig.~\ref{FigNorma}). The
spiral and irregular galaxy population appears far from relaxed 
\citep{woudt2008}. The center of the cluster is assumed to be at the 
position of the cD galaxy ESO~137-006.

\begin{table}
\centering
\caption[]
{Parameters of the ESO~137-001 galaxy.}
\label{TabPar}
\begin{tabular}{ll}
\hline
\hline
\noalign{\smallskip}
ESO~137-001 (WKK6176): & \\
\noalign{\smallskip}
\hline
\noalign{\smallskip}
RA, Dec (J2000) & $16^{\rm h}13^{\rm m}27^{\rm s}.24$, $-60^\circ45'50''.6$\\
type & SBc\\
radial velocity $\upsilon$ & $4661\pm 46$~km~s$^{-1}$\\
major/minor diameter & $1.26'$/$0.56'$\\
PA, inclination\tablenotemark{1} & $8.9^\circ$, $66.2^\circ$\\
total $B$ and $I$ magnitudes\tablenotemark{2} & $14.31\pm 0.08$, $13.20\pm 0.07$\\
stellar mass\tablenotemark{3} & $(0.5- 1.4)\times 10^{10}~M_\odot$\\
rotation velocity\tablenotemark{4} & $\sim 110$~km\,s$^{-1}$\\
luminosity distance\tablenotemark{3} & 69.6~Mpc\\
\noalign{\smallskip}
\hline
\noalign{\smallskip}
\end{tabular}
\tablecomments{ From HyperLeda catalogue \citep{paturel2003}.}
\tablenotetext{1}{\citet{sun2007} give $i=61^\circ-64^\circ$}
\tablenotetext{2}{from \citet{sun2007}, the correction for the intrinsic
extinction is not applied}
\tablenotetext{3}{from \citet{sun2010} and \citet{sivanandam2010}}
\tablenotetext{4}{from $K$-band velocity-luminosity relation 
\citep{sivanandam2010}}
\end{table}

At a projected distance of $\sim 280$~kpc from the cluster center to
the NW direction a blue galaxy ESO~137-001 is located (see
Fig.~\ref{FigNorma} and Table~\ref{TabPar}). Its line-of-sight velocity is about
-200~km~s$^{-1}$ w.r.t. the cluster mean which suggests that the main
velocity component is in the plane of the sky. $Chandra$ and
XMM-$Newton$ imaging revealed a long, narrow X-ray tail extending from
the galaxy to $\sim 80$~kpc projected distance \citep{sun2006}, see the
left panel of Fig.~\ref{FigPoints}. It points away from the direction
to the cluster center. The total estimated X-ray gas mass is about
$10^9 M_\odot$. Deeper $Chandra$ observation revealed even a fainter
secondary X-ray tail that is well separated from the primary one
\citep{sun2010}. The bright X-ray emission in the tails is presumably
arising from mixing of cold stripped ISM with the surrounding hot ICM
\citep{sun2010, tonnesen2011}. Currently, \citet{ruszkowski2014}
suggested that the tail bifurcation may arise as a consequence of
magnetized ram pressure wind that produces very filamentary morphology
of the tail.

Optical images of ESO~137-001 reveal a $\sim 40$~kpc long H$\alpha$
tail comprised of diffuse gas and discrete \ion{H}{2} regions roughly
coincident with the X-ray tail. Moreover, a number of discrete
\ion{H}{2} regions are occurring outside the visible X-ray tail,
forming a broader tail \citep{sun2007}. A large amount of star
formation corresponding to the total number of more than 30 \ion{H}{2}
regions is thus occurring well outside the main body of the galaxy.
The total mass of starbursts in the \ion{H}{2} regions in the tail is 
estimated to about $10^7 M_\odot$. 
ESO~137-001 is expected to host a significant amount of \ion{H}{1} and
H$_2$ initially. To a weak limit of $5\times 10^8 M_\odot$ no
\ion{H}{1} was detected with ATCA \citep[Australia Telescope Compact
Array,][]{vollmer2001}\footnote{The upper limit is lower than in
\citet{vollmer2001} since a lower luminosity distance of 69.6~Mpc is
adopted for the A3627 cluster. It is possible that some (extended)
\ion{H}{1} emission was filtered out by the ATCA interferometric
observation. No single-dish data are available.}. {\it Spitzer} IRS
observations however revealed more than $10^7 M_\odot$ of warm H$_2$
($\sim 160$~K) in the galaxy and inner 20~kpc of the tail
\citep{sivanandam2010}. As this warm gas is likely a small fraction of
the total molecular hydrogen, deep \ion{H}{1} and CO observations are
expected to reveal the cold component in the galaxy.

ESO~137-001 is an excellent candidate for an ongoing transformation
from a blue, gas-rich to a red, gas-poor galaxy due to violent removal
of its ISM by ram pressure stripping. It has the most dramatic tail of
a late-type galaxy observed up today, making it an ideal laboratory for
detailed studies of the complex processes that take place during ram
pressure stripping. A3627 is a rich cluster where ram pressure can be
1-2 orders of magnitude stronger than in Virgo. The high pressure
(including thermal pressure) environment might therefore be strong
enough to affect even denser (molecular) gas.
HST observations revealed a complex structure with disrupted dust
content, filaments of young stars, and bright extraplanar star clusters
in the inner tail of ESO~137-001 (see Fig.~\ref{FigPoints}, right
panel). Detailed study of these observations will be published
elsewhere.
{\it Herschel} PACS and SPIRE
imaging shows a dust trail emanating from the galaxy that is coaligned
with the warm H$_2$, H$\alpha$ and the more prominent of the two X-ray
tails (Sivanandam et al., in prep.).

In this paper we want to answer basic questions about the presence,
amount, spatial distribution, and origin of molecular gas in the
star-forming stripped tail of ESO~137-001.
With APEX, we searched for $^{12}$CO(2-1) emission in
regions covering the main body of the galaxy, as well as in the
H$\alpha$ emission peaks in the tail.
The paper is structured in the following way: observations performed
are introduced in Section~2, the results are presented in Section~3,
the efficiency of star formation in the system is studied in Section~4,
the structure of the stripped gas is analyzed in Section~5, the ram
pressure operating on the galaxy is semi-analytically estimated in 
Section~6. Discussion follows in Section~7, and the conclusions in 
Section~8. We adopt a cluster redshift of 0.0163 for A3627 
\citep{woudt2008}. Assuming $H_0= 71$~km\,s$^{-1}$\,Mpc$^{-1}$, 
$\Omega_M= 0.27$, and $\Omega_\Lambda= 0.73$, the luminosity distance 
is 69.6~Mpc, and $1''= 0.327$~kpc.

\section{Observations}\label{observations}
The observations of ESO~137-001 were carried out with the Atacama
Pathfinder EXperiment (APEX) 12\,m antenna in September
2011\footnote{In the same APEX programme, the Norma ram pressure
stripped galaxy ESO~137-002 \citep{sun2010} was also observed. The
results will be discussed in a forthcoming paper.}. Using the APEX-1
receiver of the Swedish Heterodyne Facility Instrument (SHFI) the
observations were done at the frequency of the $^{12}$CO(2--1)
($\nu_{\rm rest} = 230.538$~GHz) line. The eXtended Fast Fourier
Transform Spectrometer (XFFTS) backend was used with a total bandwidth
of 2.5~GHz divided into 32768 channels. The corresponding velocity
resolution is about 0.1~km\,s$^{-1}$. The XFFTS consists of two units
with a 1~GHz overlap region. It thus covers the entire IF bandwidth of
the SHFI. At 230~GHz the FWHM of the primary beam of the telescope is
about $25''$ which, with the adopted distance of the Norma Cluster of
$\sim 69.6$~Mpc, corresponds to about 8.4~kpc. The main
beam projected area $\Omega_B \simeq 700$~arcsec$^2= 80$~kpc$^2$,
including a factor $1/\ln2$ of a Gaussian beamshape correction. The
system temperatures were typically about 140~K and the data are of an
excellent quality.

\begin{figure*}[t!]
\centering
\includegraphics[height=0.55\textwidth]{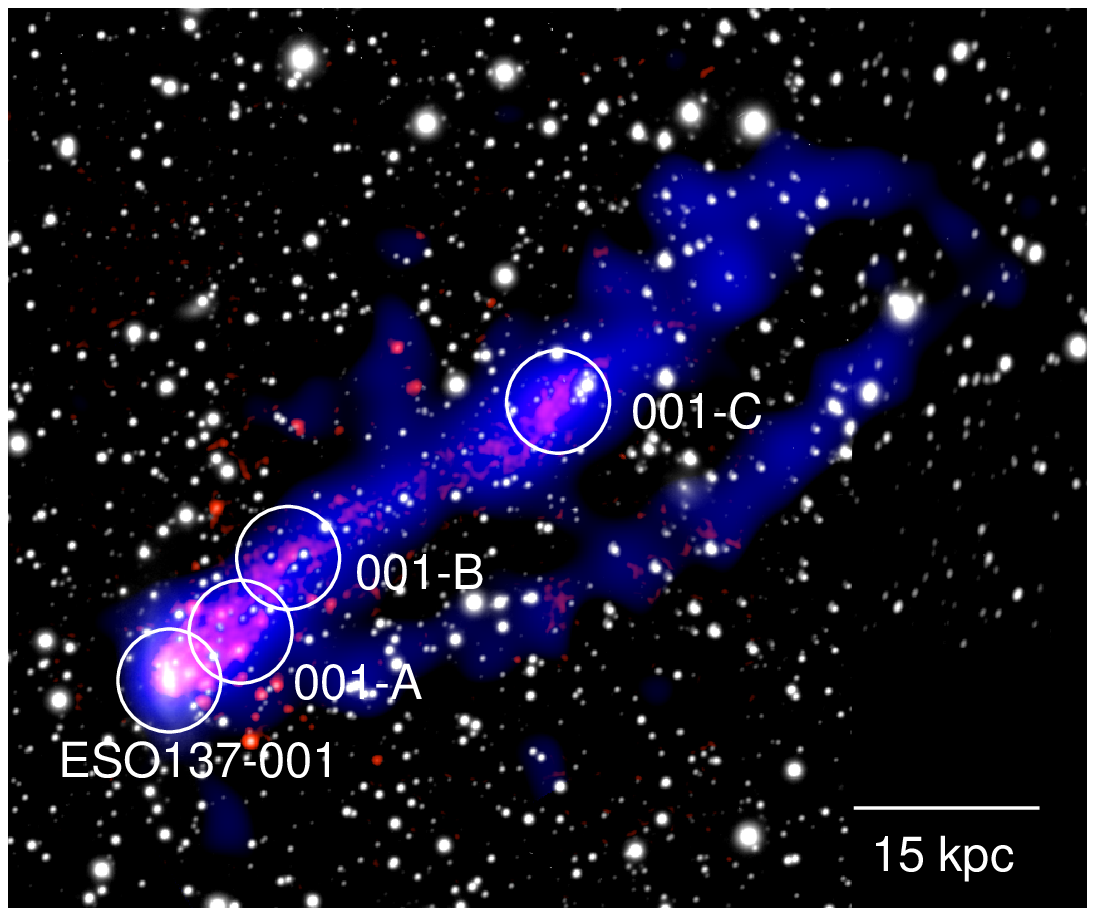}
\includegraphics[height=0.55\textwidth]{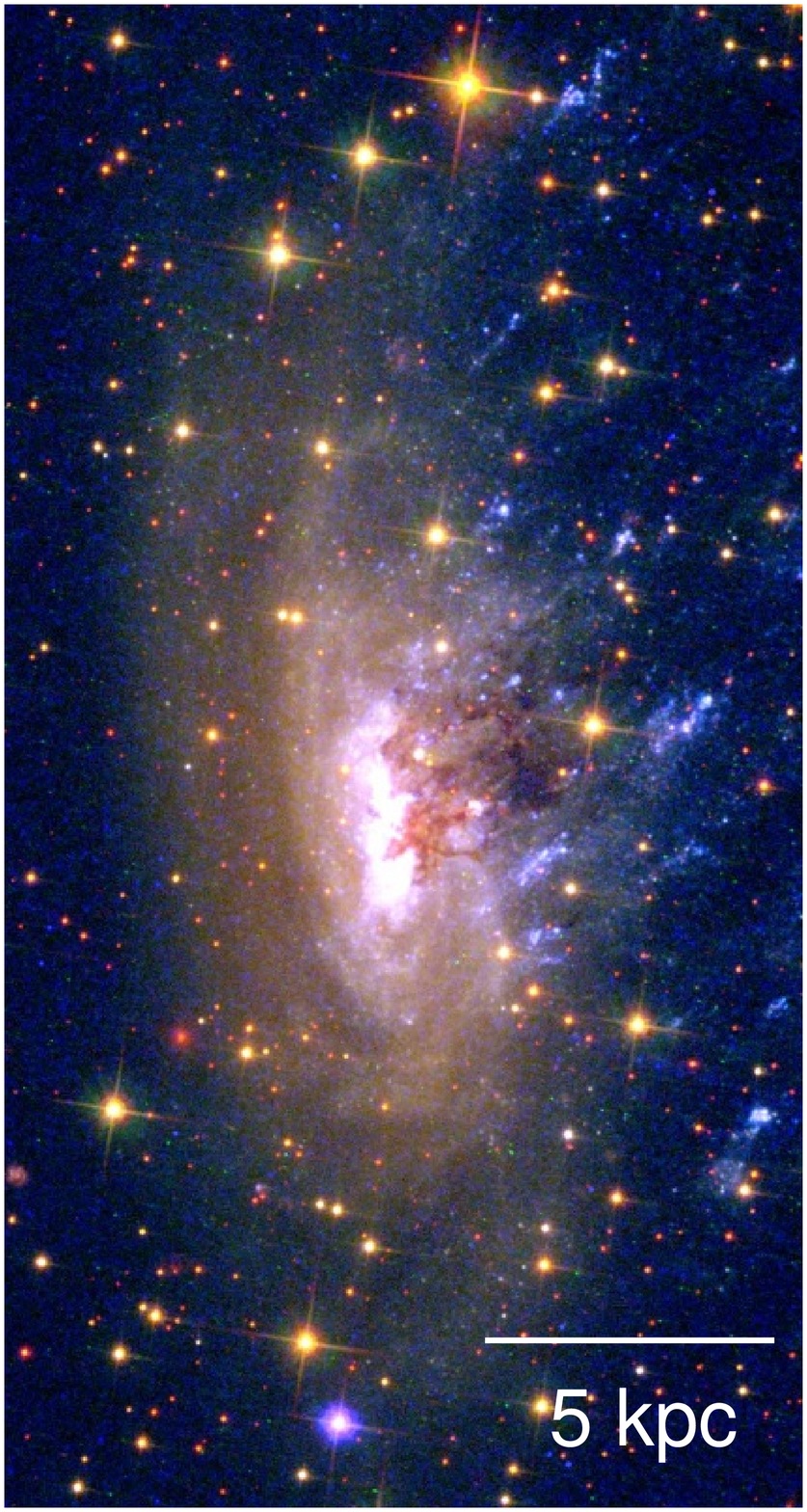}
\caption{
Left: Observed positions on a composite X-ray {\it Chandra} (blue) /
H$\alpha$ SOAR (red) image of ESO~137-001 with $^{12}$CO(2-1) beams
(${\rm FWHM}= 25''\approx 8.4$~kpc) of the APEX telescope. Nearly all
of the optical sources are foreground stars. Right: HST image of the
disk of ESO~137-001 showing the extent of the stellar component and
with a great detail the filamentary structure of extraplanar dust and
young stars. 
The 3D orientation of the
galaxy with respect to the observer is likely the following: the
observed spiral arms are trailing; the near side of the disk is to the
east (as suggested from dust extinction in the west); thus the southern
side is receding (positive radial velocities) while the northern is
approaching.
}\label{FigPoints}
\end{figure*}

\begin{table}
\centering
\caption[]{List of observed positions.}
\label{Sources}
\begin{tabular}{lcccc}
\hline
\hline
\noalign{\smallskip}
Source & R.A.    & Dec.    & $\upsilon_{\rm lsr,\,opt}$ & $T_{\rm ON}$ \\
       & (J2000) & (J2000) & (km\,s$^{-1}$)             & (min)         \\
\noalign{\smallskip}
\hline
\noalign{\smallskip}
ESO137-001& 16:13:27.24 & -60:45:50.6 & 4630 & 13  \\
001-A  & 16:13:24.75 & -60:45:39.5 & 4600 & 21  \\
001-B  & 16:13:23.30 & -60:45:21.0 & 4600 & 56  \\
001-C  & 16:13:14.34 & -60:44:42.4 & 4500 & 137 \\
\noalign{\smallskip}
\hline
\noalign{\smallskip}
\end{tabular}
\end{table}

The observations were done in a symmetric Wobbler switching mode with
the maximum throw of $300''$  in order to avoid with OFF positions the
tail if oriented in azimuth. The Wobbler switching is known to provide
better baseline stability than position switching. The integration
points in ESO~137-001 were selected to cover its main body, as well as
H$\alpha$ bright regions in its tail, including one at a large
projected distance of $\sim 2'$ from the galaxy (see scheme
in Fig.~\ref{FigPoints}). The list of observed positions is given in
Tab.~\ref{Sources}, together with information on their radial velocity
and actual on-source observing time. The receiver was tuned to the
$^{12}$CO(2-1) frequency shifted in each position to its respective
optical radial velocity (given in Table~\ref{Sources}).
Observing conditions were good with PWV less than 1~mm.

The data were reduced according to the standard procedure using CLASS
from the GILDAS software package developed at IRAM. Bad scans were
flagged and emission line-free channels in the total width of about
1000~km\,s$^{-1}$ were used to subtract first-order baselines.
The corrected antenna temperatures, $T^*_{\rm A}$, provided by the APEX
calibration pipeline \citep{dumke2010}, were converted to
main-beam brightness temperature by $T_{\rm mb}=T^*_{\rm A} / \eta_{\rm 
mb}$, using a main beam efficiency of about $\eta_{\rm mb} =0.75$ at
230~GHz. The rms noise levels typically of $1-2$~mK per
12.7~km~s$^{-1}$ channels were obtained. Gaussian fits were used to
measure peak $T_{\rm mb}$, width, and position of the detected CO
lines. Flux densities can be obtained using the conversion factor
$S_\nu/T_{mb}=39$~Jy\,beam$^{-1 }$\,K$^{-1}$ for the APEX telescope.

\section{Results}\label{results001}
\begin{figure*}[t]
\centering
\includegraphics[height=0.48\textwidth,angle=270]{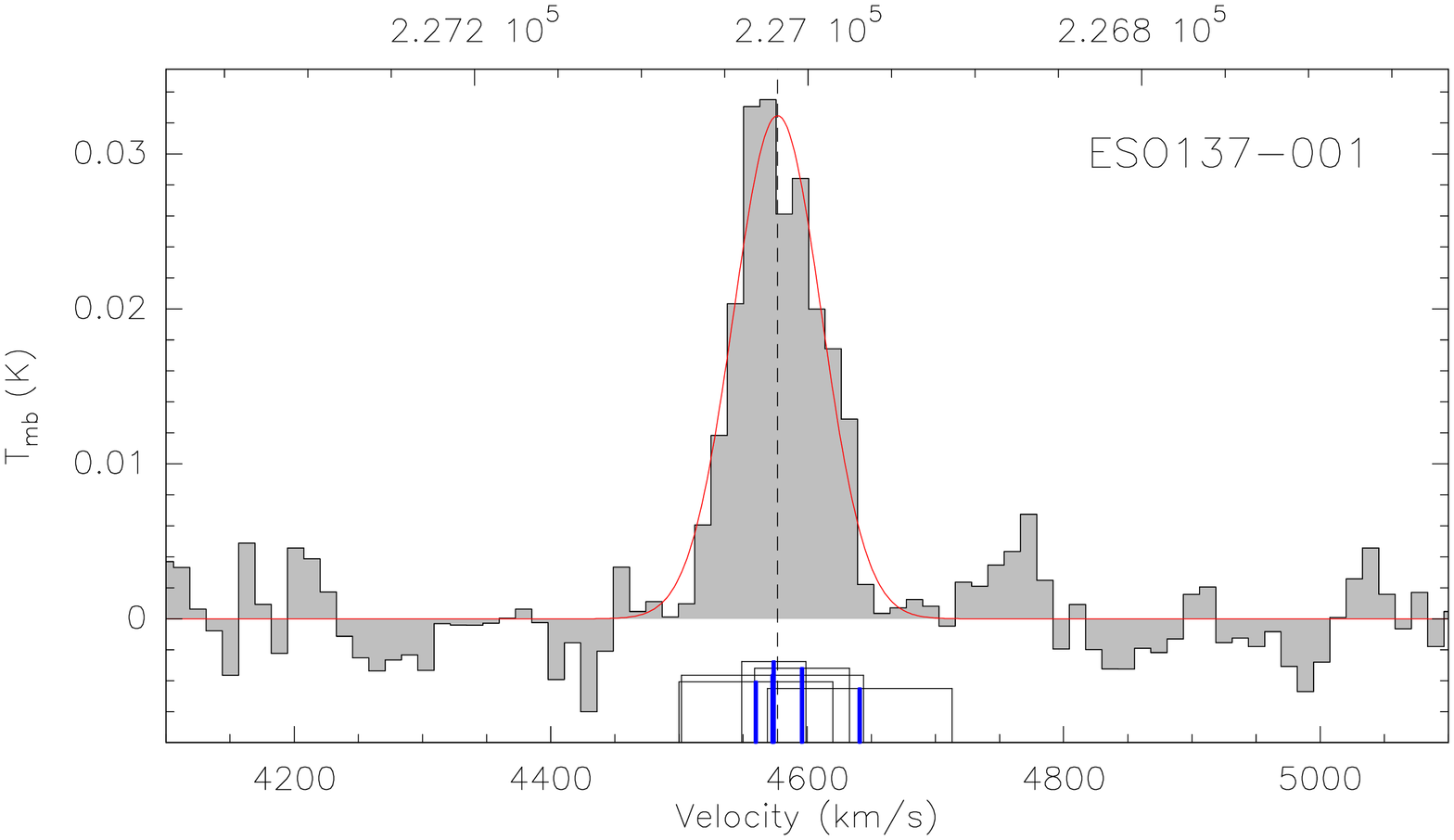}
\hspace{0.2cm}
\includegraphics[height=0.48\textwidth,angle=270]{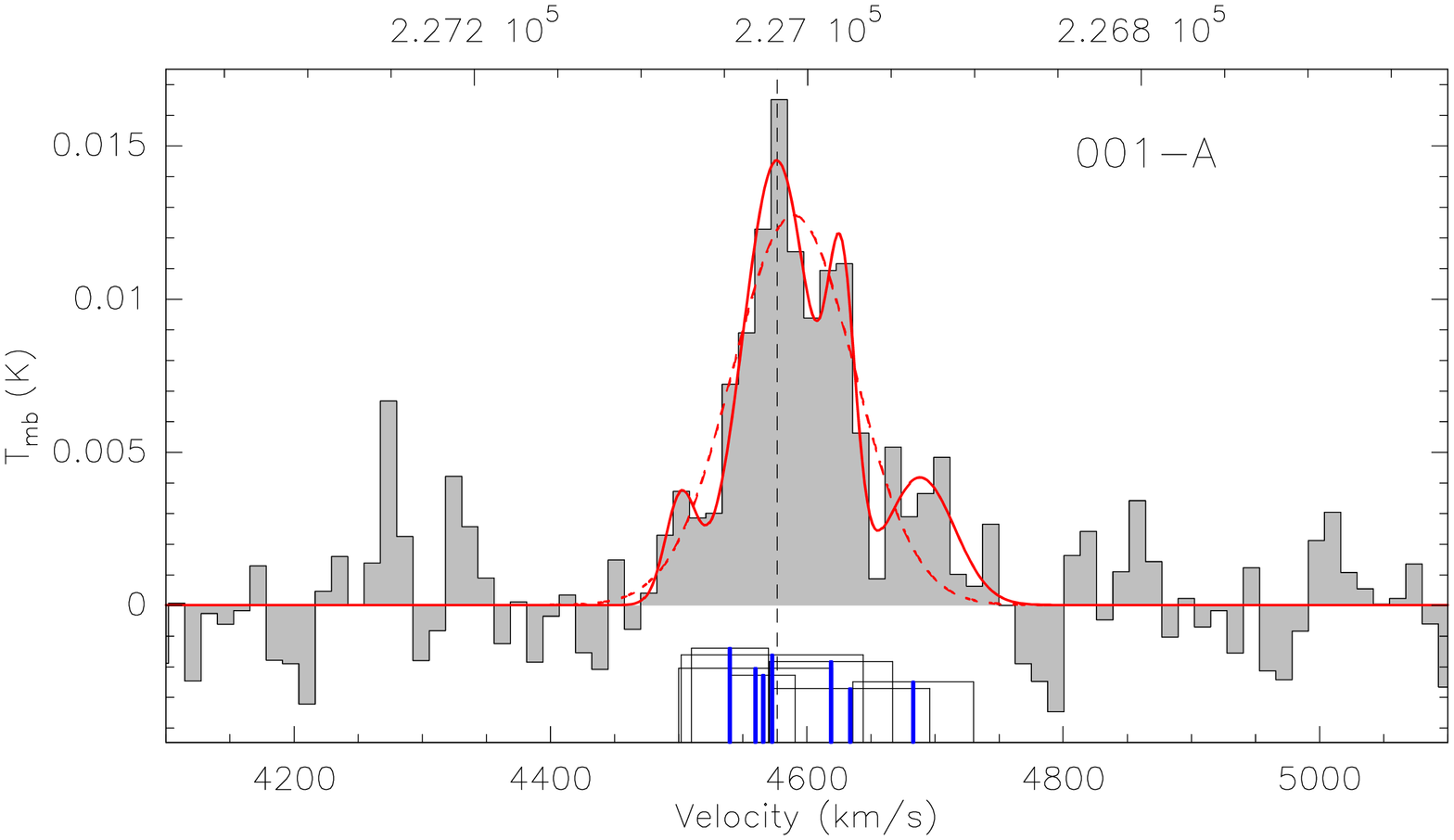}
\vspace{0.4cm}\\
\includegraphics[height=0.48\textwidth,angle=270]{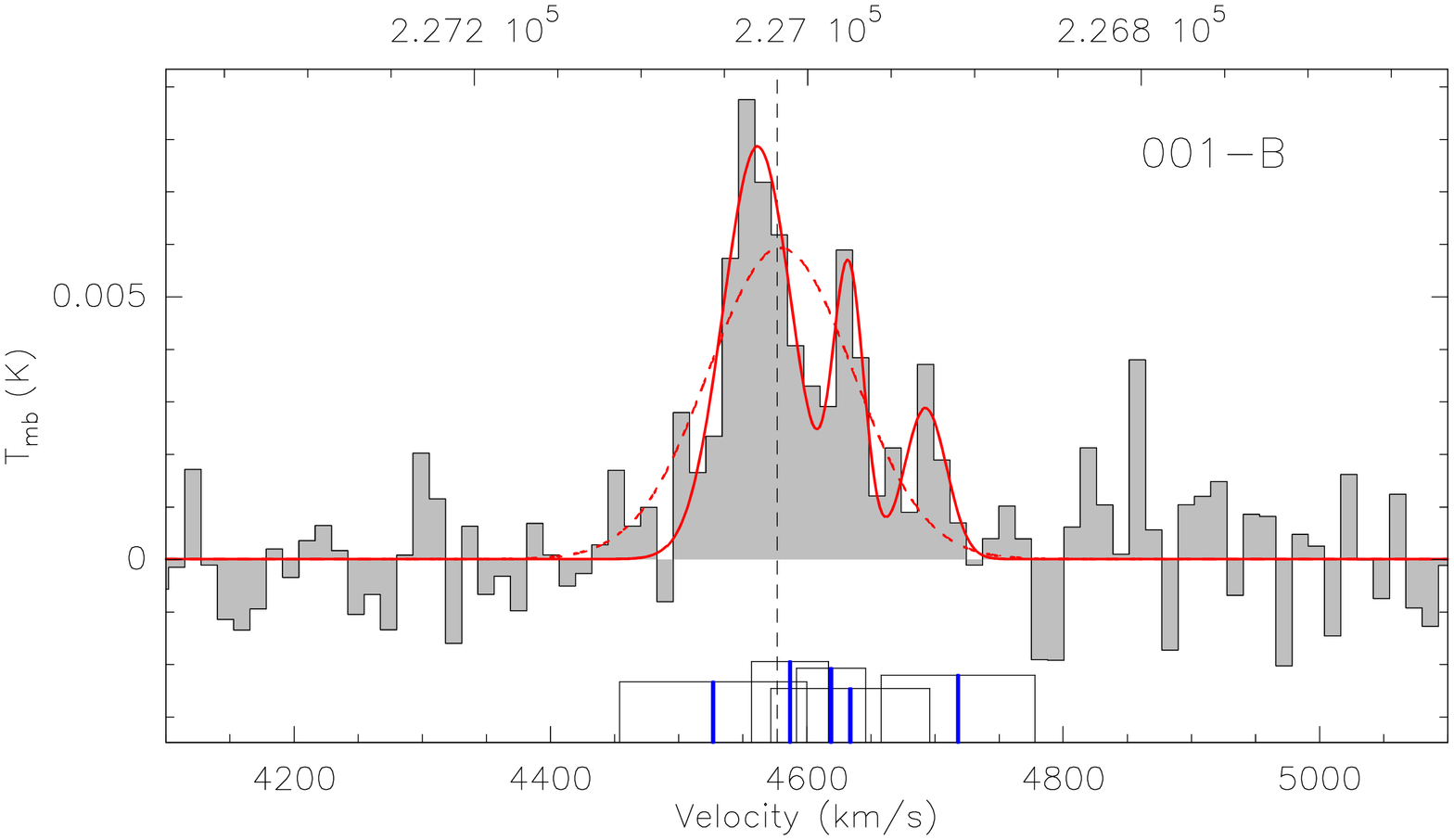}
\hspace{0.2cm}
\includegraphics[height=0.48\textwidth,angle=270]{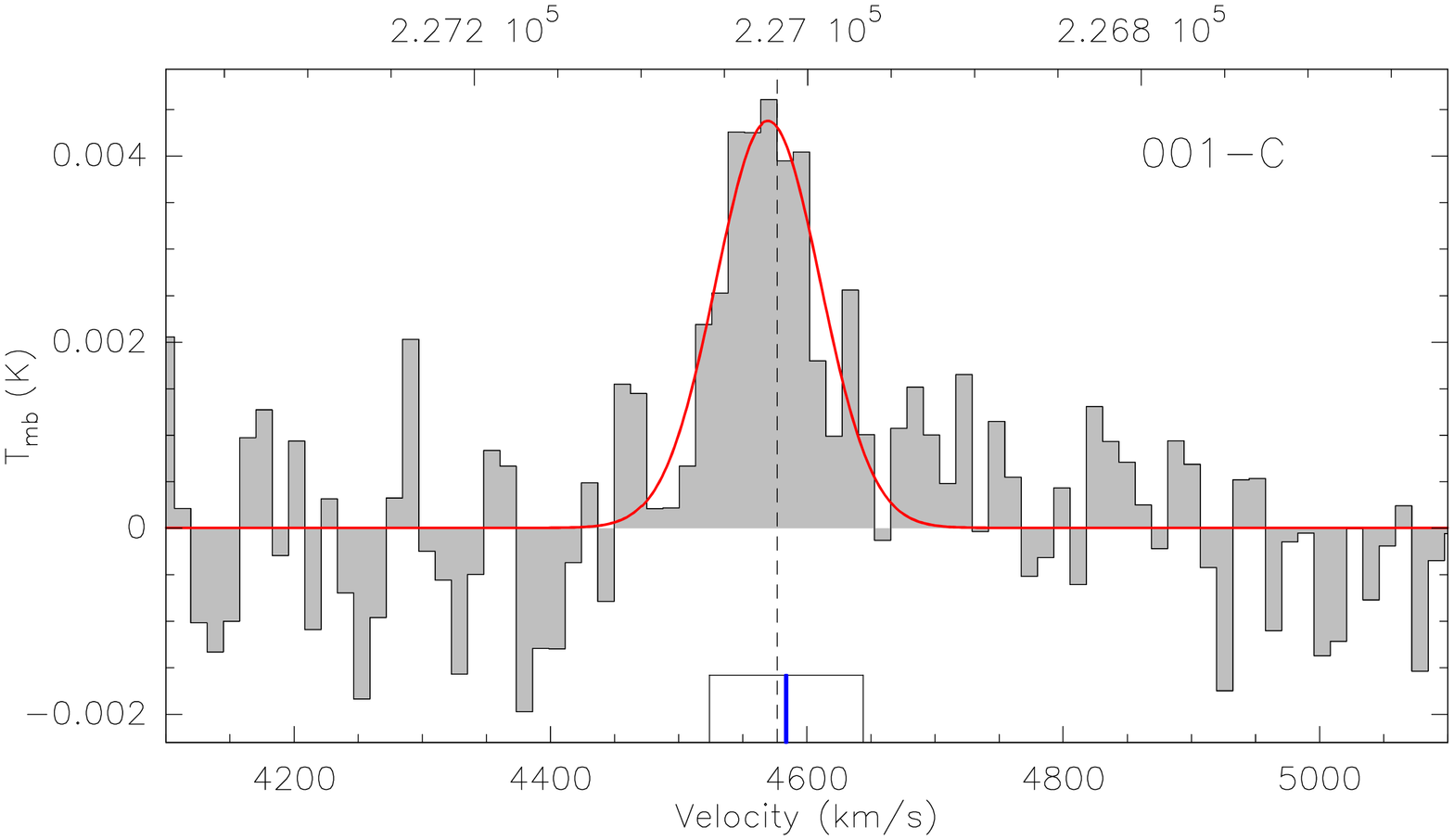}
\caption{
The CO(2-1) emission spectra of ESO~137-001 in the main body pointing
(top left), the off-disk positions 001-A (top right) and 001-B (bottom
left) in the inner tail, and the 001-C position in the outer tail.
Spectra are smoothed to 12.7~km\,s$^{-1}$ channels. Parameters of the
Gaussian fits to the lines are given in Table~\ref{TabRes}. The central
velocity 4577~km\,s$^{-1}$ of the fit is shown with the dashed line.
The radial velocities of the \ion{H}{2} regions encompassed by the
APEX beam in each pointing are marked with the blue ticks with wings
corresponding to the velocity errors \citep{sun2007}. Their heights
order the regions according to H$\alpha$ luminosity.
}\label{Fig001}
\end{figure*}

\subsection{Main body of ESO 137-001}
In the pointing centered on the main body of ESO~137-001, CO(2-1)
emission is strongly detected (see Fig.~\ref{Fig001}, the top left
panel). The corresponding molecular gas mass is about $1.1\times 10^9 
M_\odot$, following $M_{\rm H_2}\, [M_\odot]= 5.5\, L_{\rm CO(2-1)}\, 
[{\rm K\,km\,s^{-1}\,pc^2}]$, where we assume a typical values for the
CO(2-1)-to-CO(1-0) ratio of 0.8 \citep[e.g., ][]{leroy2009}, and a
standard Galactic CO(1-0)-to-H$_2$ conversion factor of $2\times 
10^{20}$~cm$^{-2}$(K\,km\,s$^{-1})^{-1}$ \citep[e.g.,
][]{kennicutt2012, feldmann2012}. The formula includes a factor of 1.36
to account for the presence of Helium. Using a Gaussian fit (plotted in
Fig.~\ref{Fig001} over the spectrum) we measure the linewidth of
81~km\,s$^{-1}$ and the central velocity of the line of $\sim 
4577$~km\,s$^{-1}$, assuming the radio definition of the
velocity\footnote{
$\upsilon_{\rm rad}= c\,(1- \nu_{\rm sky}/ \nu_{\rm 0})\approx 
4577$~km\,s$^{-1}$}.
Other parameters of the line are given in Table~\ref{TabRes}.
\citet{sun2007} measured the radial velocity of the central emission
nebula from two different optical CTIO~1.5~m spectra to $4667\pm 
135$~km\,s$^{-1}$ and $4640\pm 20$~km\,s$^{-1}$, which is very close to
our measurement\footnote{
$\upsilon_{\rm opt}= c\,(\nu_{\rm 0}/ \nu_{\rm sky} -1)\approx 
4648$~km\,s$^{-1}$.}.

As is apparent from Fig.~\ref{FigPoints}, the central APEX beam has
been contaminated by a contribution from off-disk regions. Within its
$\sim 8$~kpc FWHM it encompasses not only the central H$\alpha$
emission nebula but also five identified off-disk \ion{H}{2} regions
\citep{sun2007}, including the brightest one \citep[ELO1 in][]{sun2007}
that is located only about 2~kpc downstream in the tail. Their radial
velocities are mostly close ($\lesssim 30$~km\,s$^{-1}$) to the
galaxy's system velocity, except one at $54\pm 72$~km\,s$^{-1}$ located 
near the southern beam edge. The radial velocities of the encompassed 
\ion{H}{2} regions are shown at the bottom of the spectrum panel in 
Fig.~\ref{Fig001} with the blue bars with black wings corresponding to 
the errors. The ratio of H$\alpha$ luminosities of the central nebula 
and of the brightest off-disk region ELO1 is $\sim 10$ \citep{sun2007}. 
This suggests that the central nebula could dominate also the CO 
emission in the main body pointing, however, we don't know what is the 
relation between H$\alpha$ and CO emission and thus cannot say what 
fraction of the total H$_2$ mass of $\sim 1.1\times 10^9 M_\odot$ 
detected in the main body point belongs to the off-disk sources.

\begin{table*}[t]
\centering
\caption[]
{Properties of the detected CO(2-1) lines.}
\label{TabRes}
\begin{tabular}{lccccccc}
\hline
\hline
\noalign{\smallskip}
Source & rms & velocity & FWHM & $T_{\rm 
peak}$ & $I_{\rm CO, fit}$ & $I_{\rm CO}$ & $M_{\rm mol}$\\
       & (mK) & (km\,s$^{-1}$) & (km\,s$^{-1}$) & (mK) & 
(K~km~s$^{-1}$) & (K~km~s$^{-1}$) & ($10^8 M_\odot$)\\
\noalign{\smallskip}
\hline
\noalign{\smallskip}
ESO137-001 & 2.6 & $4576.9\pm 1.8$  &   $80.7\pm 4.0$ & 32.5 & $2.8\pm 0.1$   & 2.7 & $11.1$\\
\hline\noalign{\smallskip}
001-A     & 1.9 & $4589.6\pm 4.0$  & $111.8\pm 10.9$ & 12.7 & $1.5\pm 0.1$   & 1.6 & $6.4$\\
\cline{2-8}\noalign{\smallskip}
          & & $4501.4\pm 8.7$  & $28.6\pm 21.3$ &  3.5 & $0.11\pm 0.07$ & - & $0.5$\\
          & & $4576.3\pm 3.7$  & $62.5\pm 12.5$ & 14.5 & $0.97\pm 0.15$ & - & $4.0$\\
          & & $4626.7\pm 3.6$  & $25.7\pm  7.7$ &  9.3 & $0.25\pm 0.11$ & - & $1.0$\\
          & & $4688.3\pm 9.9$  & $63.2\pm 22.2$ &  4.2 & $0.28\pm 0.09$ & - & $1.2$\\
\hline\noalign{\smallskip}
001-B     & 1.2 & $4579.4\pm 6.3$  & $132.5\pm 16.8$ &  5.9 & $0.84\pm 0.08$ & 0.8 & $3.4$\\
\cline{2-8}\noalign{\smallskip}
          & & $4561.3\pm 3.4$  & $63.4\pm 11.2$ &  7.9 & $0.53\pm 0.07$ & - & $2.2$\\
          & & $4632.2\pm 4.2$  & $28.4\pm 14.8$ &  5.5 & $0.17\pm 0.07$ & - & $0.7$\\
          & & $4692.3\pm 8.7$  & $39.0\pm 29.7$ &  2.9 & $0.12\pm 0.06$ & - & $0.5$\\
\hline\noalign{\smallskip}
001-C     & 1.0 & $4569.5\pm 5.5$  &  $95.9\pm 17.0$ &  4.4 & $0.45\pm 0.06$ & 0.4 & $1.5-1.8$\\
\hline
\end{tabular}
\tablecomments{
The table gives the $1\sigma$ rms in 12.7~km\,s$^{-1}$ channels,
parameters of (multiple) Gaussian fits (the line heliocentric central 
velocity, the FWHM, the peak temperature, and the integrated 
intensity), the measured integrated intensity, and the molecular
gas mass. The temperatures are given in $T_{\rm mb}$ scale. First order
baselines were subtracted in the velocity range $4100 - 
5100$~km\,s$^{-1}$. Radio definition of velocity was used to convert
the sky frequency of the source.}
\end{table*}

\subsection{Molecular gas in the halo -- the inner tail of ESO~137-001}
In the innermost tail pointing 001-A, located roughly one beam ($\sim 
8$~kpc) downstream of the galaxy center, CO(2-1) emission is detected
with very high significance (see Fig.~\ref{Fig001}, the top right
panel). The measured CO luminosity corresponds to $6.4\times 10^8 
M_\odot$ of molecular gas, about half the mass found in the main body
position. Although the X-factor is uncertain for the special
environment of a gas stripped tail, we assume the Galactic value
because the metallicity measured in the tail is close to solar \citep[$\sim 
0.6~Z_\odot$,][]{sun2010}. The actual value of the X-factor
will be discussed in a forthcoming paper based on observations of
other CO transitions in the system.
Multiple components are visible in the 001-A line -- a wing at the
low-velocity side, a double-profile main part, and a $4\sigma$ feature 
at the high-velocity side centered at $\sim 4687$~km\,s$^{-1}$. In
Table~\ref{TabRes} the parameters of the multiple Gaussian fits are
given together with a single Gaussian fit of the whole line. The
central velocity of the single fit is only slightly higher ($\sim 
4590$~km\,s$^{-1}$) than in the main body position. We discuss the
components in more detail in Subsection~\ref{multiple}.
The 001-A pointing encompasses three bright \ion{H}{2} regions at
radial velocities $-50\pm 30$, $-23\pm 25$, and $31\pm 48$~km\,s$^{-1}$
with respect to the galaxy's systemic velocity,
as well as several weaker regions including one at the edge of the beam
that has a higher radial velocity ($97\pm 47$~km\,s$^{-1}$) which
corresponds well to the high-velocity $4\sigma$ feature in the CO
spectrum.

Advancing another $\sim 8$~kpc along the tail to the 001-B point, about
$3.4\times 10^8 M_\odot$ of molecular gas is detected (see
Fig.~\ref{Fig001}, the bottom left panel), about half of the amount
found in 001-A. The line profile again shows several velocity
components that are more distinct than in the case of 001-A. The
parameters of a multiple Gaussian profile fit are given in
Table~\ref{TabRes}. Although not included in the fitting, there is a
hint of a low-velocity feature similar to that in 001-A. The 001-B
pointing encompasses several bright \ion{H}{2} regions which are
moving at about $-64$~km\,s$^{-1}$ to $+133$~km\,s$^{-1}$ relative to
the systemic
velocity along the line of sight. Thus CO and H$\alpha$ velocities
are again well consistent.

In total, in the inner tail regions 001-A and 001-B we thus detected
almost the same amount ($\sim 90\%$) of molecular gas as in the galaxy
pointing. Taking into account the contribution of off-disk regions
to the main body point intensity, and the uncovered parts of the tail,
there probably is more H$_2$ in the inner $\sim 20$~kpc of the tail
than in the main body.

\subsection{IC molecular gas -- the outer tail of ESO~137-001}
Strong CO(2-1) emission is detected even in the outermost position
we have searched located at $\sim 40$~kpc (in projection) from the main
body, where both X-ray and H$\alpha$ emission peak (see 
Fig.~\ref{Fig001}, the bottom right panel).
This is the first time CO emitting gas is detected in intra-cluster
regions of a clearly ram pressure stripped tail, that moreover is
associated spatially with star-forming regions and X-ray emitting gas.
The corresponding molecular gas mass is $\sim 1.8\times 10^8 M_\odot$,
which is about a factor 2 less than in the region 001-B, and it
constitutes about 15\% of the total H$_2$ mass found in the tail. 
As we will discuss later in Section~\ref{gradient}, higher molecular 
gas temperature in the tail (as well as optical thickness and other 
effects), especially at the farthest pointing 001-C, could possibly 
decrease the CO-to-H$_2$ conversion factor from a Galactic value that 
we used, and therefore yield an overestimate of the H$_2$ content 
\citep{bolatto2013}. 

A single Gaussian fit of the CO line is centered at $\sim 4570$~km\,s$^{-1}$
which is very close to the central velocity of the main body line.
This indicates that indeed the gas-stripped tail extends almost
entirely in the plane of the sky and suggests that also the orbital
motion of the galaxy happens mostly in the plane of the sky.
Table~\ref{TabRes} further summarizes the parameters of the line fit.
The CO line profile is asymmetric, indicating again an internal
structure.
Only one \ion{H}{2} region identified by \citet{sun2007} occurs in the
001-C pointing area. Its radial velocity is well consistent with the
central velocity of the CO line (see the blue bar in the bottom right
panel of Fig.~\ref{Fig001}).

The tidal truncation radius of the dark matter halo of ESO~137-001 is
estimated to be about 15~kpc \citep{sun2010}. This suggests that the
molecular gas detected in the 001-C region could already be released
from the gravitational well of the galaxy and thus contribute directly
to the intracluster material in the Norma cluster. Due to the lack of
(almost) any radial velocity component however, we cannot determine
whether the detected molecular gas is indeed at velocities exceeding
the escape speed from the galaxy or stays bound to it.

\section{Efficiency of star formation}
The optical spectroscopy and the H$\alpha$ imaging revealed star-forming 
nucleus of ESO~137-001, as well as over 30 \ion{H}{2} regions 
downstream in the tail \citep{sun2007, sun2010}. Subsequent infrared 
data showed associated $8~\mu$m excess sources in the inner $\sim 1'$ 
of the tail \citep{sivanandam2010}. In this section, we estimate how 
efficient has been the transformation of the detected molecular gas 
into stars. We are especially interested in the situation in the 
gas-stripped tail of the galaxy. 

\begin{table}[t]
\centering
\caption[]
{The H$\alpha$-based SFRs in the observed regions.
}
\label{SF}
\begin{tabular}{lccc}
\hline
\hline
\noalign{\smallskip}
Source & $L_{\rm H\alpha}$ & SFR$_{\rm H\alpha}$ & $L_{\rm CO}$\\
       & (erg\,s$^{-1}$)   & ($M_\odot$\,yr$^{-1}$) & (K\,km\,s$^{-1}$\,pc$^2$)\\
\noalign{\smallskip}
\hline
\noalign{\smallskip}
ESO137-001& $6.7\times 10^{40}$ & $3.6\times 10^{-1}$ & $2.0\times 10^8$\\
001-A     & $6.3\times 10^{39}$ & $3.4\times 10^{-2}$ & $1.1\times 10^8$\\
001-B     & $1.4\times 10^{39}$ & $7.5\times 10^{-3}$ & $5.0\times 10^7$\\
001-C     & $1.7\times 10^{38}$ & $9.2\times 10^{-4}$ & $4.0\times 10^7$\\
\noalign{\smallskip}
\hline
\noalign{\smallskip}
\end{tabular}
\tablecomments{The H$\alpha$ luminosities from \ion{H}{2} regions covered with our 
APEX pointings (circles with a diameter of $25''$) and corresponding star
formation rates, together with measured CO luminosity.}
\end{table}

\subsection{Star formation in the disk}
The rate of ongoing star formation (SFR) corresponding to the H$\alpha$
luminosity of the main body of ESO~137-001 can be calculated from the
\citet{kennicutt2012} relation
\begin{equation}\label{kennicutt}
\log{\rm SFR}\ (M_\odot\,{\rm yr}^{-1})= \log{L_{\rm H\alpha}}\ ({\rm 
erg\,s^{-1}}) - 41.27
\end{equation}
that uses the initial mass function (IMF) of \citet{kroupa2003}.
The H$\alpha$ luminosity of \ion{H}{2} regions covered by the
ESO~137-001 main body APEX pointing is $\sim 6.7\times 
10^{40}$~erg\,s$^{-1}$ (see Table~\ref{SF}), dominated by the central
nebula. The corresponding SFR is $\sim 0.4~M_\odot$\,yr$^{-1}$. It has
been shown that multiwavelength observations may reduce uncertainties
and errors in SFR determination caused mainly by dust attenuation,
especially for edge-on galaxies. While there is no data for ESO~137-001 
from Spitzer/MIPS, the \textit{Wide-Field Infrared Survey Explorer} 
(WISE) measured in its band 4 (22~$\mu$m) a flux density of 58.7~mJy. 
Using the SFR WISE calibration relations of \citet{lee2013} corrected 
for Kroupa IMF, in combination with the H$\alpha$ flux of the galaxy
\citep{sun2007}, the corresponding SFR of the main body of ESO~137-001
is $\sim 1~M_\odot$\,yr$^{-1}$.

\begin{figure}[t]
\centering
\includegraphics[width=0.47\textwidth]{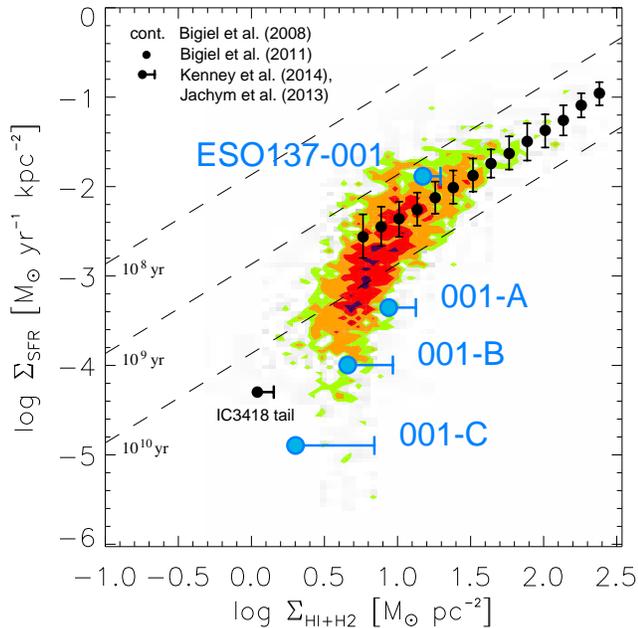}
\caption{
Star formation rate surface density as a function of total gas surface
density in ESO~137-001 and the three observed tail regions. The filled
circles correspond to the molecular gas column density, while the
'error bars' show ATCA \ion{H}{1} upper limits. For comparison,
measurements from sub-kpc sampling of disks of nearby spiral galaxies
\citep[][]{bigiel2008} are shown, with $\Sigma_{\rm SFR}$ inferred from
H$\alpha$ emission. The contours correspond to pixel-by-pixel sampling
of \citet{bigiel2008} of the optical disks of seven nearby spirals:
contours of 1, 2, 5, and 10 sampling points per cell (0.05~dex size)
correspond to green, orange, red, and magenta colors, respectively. The
black dots show average molecular gas depletion time measured in 30
nearby galaxies \citep[IRAM HERACLES CO survey,][]{bigiel2011}.
For comparison, also the situation in the outer parts of the
star-forming tail of the Virgo IC3418 dwarf galaxy where \ion{H}{1} was
detected and H$_2$ measured to upper limits \citep{kenney2014, 
jachym2013}
is shown. Figure adapted from \citet{bigiel2008}.
}\label{FigBigiel2008}
\end{figure}

In Fig.~\ref{FigBigiel2008}, the deduced star formation rate in
ESO~137-001 averaged over the observed APEX 230~GHz beam area ($\sim 
80$~kpc$^2$) is plotted into a
Kennicutt-Schmidt-type plot showing the $\Sigma_{\rm SFR}$ as a
function of the total (molecular + atomic) gas column density. In the
plot, measurements from disks of seven nearby spiral galaxies sampled
at sub-kpc scales by \citet[][their Fig.~8]{bigiel2008} are shown for
comparison. The filled circle at the ESO~137-001's position in
Fig.~\ref{FigBigiel2008} correspond to the H$_2$ column density, while
the ATCA \ion{H}{1} upper limit is displayed with the horizontal error
bar. Thus, in the main body of ESO~137-001, the measured molecular gas
depletion time (i.e., consumption timescale due to star formation, also
known as the inverse of the star formation efficiency) is well
consistent with average values found in other nearby spiral galaxies
\citep[$\tau_{\rm dep, H_2}\approx 2.35$~Gyr,][see black dots in
Fig.~\ref{FigBigiel2008}]{bigiel2011}.
The (total) gas surface density of $\sim 9~M_\odot$\,pc$^{-2}$ was
identified as a limit below which the star formation efficiency starts
to vary substantially (at a given gas surface density). 
\citet{bigiel2008} propose that this is due to local conditions, such 
as metallicity, gas pressure, etc., that start to play a role. In such 
places, $\Sigma_{\rm HI} > \Sigma_{\rm H_2}$ typically. The averaged 
H$_2$ column density in the main body ESO~137-001 pointing exceeds this 
limit.

\subsection{Star formation in the tail}\label{sfr}
In the gas-stripped tail of ESO~137-001, the only observed SF tracer in 
all our APEX regions is 
H$\alpha$ emission that gives evidence of recent star formation over 
the last $\sim 10$~Myr\footnote{We note that the classical 
SFR-luminosity relation is expected to work well for (a) continuous 
star formation approximation where SFR is constant over the timescale 
of H$\alpha$ emission and (b) regions where H$\alpha$ emission averages 
over more emitting sources. We are not sure how well the two conditions 
are fulfilled in the ESO~137-001 tail. The estimate of SFRs in the 
complex environments of a gas-stripped tail is thus far from trivial.}. 
Table~\ref{SF} summarizes H$\alpha$ luminosities of \ion{H}{2} regions 
covered by our APEX pointings (circles with a $25''$ diameter). 
In Table~\ref{SF}, a 1~mag extinction correction was applied, which is
probably too high for the tail regions, however, this only makes
conclusions of this section stronger. Following Eq.~\ref{kennicutt},
the SFR in the innermost tail region 001-A is low, $\sim 
0.03~M_\odot$\,yr$^{-1}$. It further drops by a factor of 4.5 when
going to the region 001-B, and by another factor of $\sim 8$ in the
most distant IC position 001-C, down to $\sim 
10^{-3}~M_\odot$\,yr$^{-1}$ (see Table~\ref{SF}). The total star 
formation rate in the three tail regions is $\sim 
0.04~M_\odot$\,yr$^{-1}$. This amount is consistent with predictions of 
numerical simulations by \citet{tonnesen2012} who found the total SFR 
saturated at $\sim 0.05~M_\odot$\,yr$^{-1}$ in the ram pressure 
stripped tail of their (massive) model galaxy. 
The low levels of star formation in the inner tail are also suggested 
from \textit{Spitzer}/MIPS $24~\mu$m-band observations of 
\citet{sivanandam2010} who measured in a region that covers partially 
our 001-A and 001-B apertures a flux of less than 10\% of the value 
from their nuclear region that roughly corresponds to our ESO~137-001 
aperture. 

The H$\alpha$-based star formation rates averaged over the APEX beams
are for the observed tail pointings also placed into the
Kennicutt-Schmidt-type plot in Fig.~\ref{FigBigiel2008}. Again, the
filled circles correspond to H$_2$ column densities, while the ATCA
upper limits on the \ion{H}{1} mass per pointing ($\sim 
4.7~M_\odot$\,pc$^{-2}$) are taken into account and displayed with the
horizontal error bars. 
Clearly, the location of the points 001-A, 001-B, and 001-C in
Fig.~\ref{FigBigiel2008} indicates much lower efficiency of star
formation in the gas-stripped tail of ESO~137-001 than is typical in
star-forming ISM in the inner parts of spiral disks. The molecular gas
depletion times are above $10^{10}$~yr, thus exceeding a Hubble time,
which indicates that most of the stripped gas does not form stars but
remains gaseous and ultimately joins the ICM. Similarly low SFEs are
found in the outer disks of spiral galaxies, in \ion{H}{1}-rich dwarf irregular
galaxies \citep[e.g.,][]{huang2012, lee2009}, or in low surface
brightness disks \citep[e.g.,][]{wyder2009}, where however, \ion{H}{1}
is likely to represent most of the ISM and CO is mostly undetected
\citep{bigiel2010}.
The CO-bright gas-stripped tail of
ESO~137-001 (with only upper limits on the \ion{H}{1} content) thus
represents a special environment of SF distinct from the above typical
categories.

While the Kennicutt-Schmidt law finds a tight correlation between the 
surface densities $\Sigma_{\rm SFR}$ and $\Sigma_{\rm gas}$ in galactic 
disks for which the typical unobserved scale height is $\sim 100$~pc, 
in the tail of ESO~137-001 the unobserved thickness is likely a few kpc. A low 
average density of the gas that is distributed in a wide 3D tail is 
thus probably an important factor in the observed low SFE. Another 
factor could be ram pressure induced shocks that may increase thermal 
and turbulent pressure of the stripped ISM \citep{sivanandam2010}. 
Turbulence can suppress star formation by inhibiting collapse of 
molecular clouds. 

In Fig.~\ref{FigBigiel2008} we also plot for comparison the total gas
column density (\ion{H}{1} detected, H$_2$ measured to upper limits)
averaged over the outer parts ($\sim 35$~kpc$^2$) of the star-forming
tail of the Virgo cluster ram pressure stripped dwarf galaxy IC3418,
where the total H$\alpha$ luminosity of the \ion{H}{2} regions
corresponds to a star formation rate of $\sim 1.9\times 
10^{-3}~M_\odot$\,yr$^{-1}$ \citep{kenney2014, jachym2013}. The
resulting star formation efficiency is again lower than $10^{10}$~yr. 

Similarly, in the ram pressure stripped extraplanar gas of several 
Virgo cluster spirals, \citet{vollmer2012} and \citet{boissier2012} 
found a depressed SFE with respect to the available amount of total 
gas. In one case (NGC~4438), \citet{vollmer2012} determined a lower SFE 
even with respect to the molecular gas. They suggested that the 
stripped gas loses the gravitational confinement and associated 
pressure of the galactic disk
which leads to a decrease in gas density and consequently in star 
formation. 
As we will discuss later in Section~\ref{pressure}, the thermal ICM 
pressure at the location of ESO~137-001 in the Norma cluster however is 
estimated to be similar to midplane pressures in galactic disks. 
Low SFEs are known also from other H$_2$ luminous sources,
such as radio galaxies where radio-loud AGNs are assumed to enhance
heating and turbulence in the molecular gas in disks, or e.g., in the
Stephan's Quintet where turbulence invoked by galaxy-tidal arm
collision is a probable source of heating \citep[e.g.,][]{nesvadba2010,
guillard2012}.

\subsection{Ram Pressure Dwarf Galaxy in the making?}
With the detected large reservoir of $>10^8~M_\odot$ of molecular gas, 
the (intra-cluster) region 001-C resembles tidal dwarf galaxies 
\citep[TDGs; e.g.,][]{braine2001} in that it is young, it also has 
formed by condensation of pre-enriched matter that belonged to a parent 
galaxy, it is now (probably) decoupled, and it may be gravitationally 
bound. We thus speculate it could be an example of a ram pressure dwarf 
galaxy (aka RPDG) in the making from the local collapse of abundant 
(stripped) gas. While in TDGs a typical molecular gas fraction is $\sim 
20\%$ (and it may grow as more \ion{H}{1} transforms into H$_2$), in 
001-C the dominant gas phase is likely H$_2$. 

However, in the extreme environment of the outer gas-stripped tail 
embedded in a hot ICM, the actual amount of molecular gas in the 001-C 
region may be overestimated by a galactic X-factor. Moreover, while in 
TDGs the efficiency of star formation was measured to be similar to 
spirals, it is much lower in the 001-C pointing. Also, the suggested 
value of star formation rate in the 001-C region (of the order of 
$10^{-3}~M_\odot$\,yr$^{-1}$) is similar to values found in compact 
knots of collisional debris of interacting galaxies rather than in TDGs 
\citep{duc2013}.
However, it could be that in 001-C we are only observing the onset of 
star formation. 
Thus, to decide whether within the 001-C pointing we are observing only 
a compact star-forming knot, or (for the first time) a ram pressure 
stripped dwarf galaxy building up, more detailed observations, 
including sensitive \ion{H}{1} search, are needed to determine the 
total mass of the object, its kinematics, and especially its 
self-gravitation.

\section{Gas phases in the stripped tails}\label{current}
In order to better understand the structure of the gas-stripped tail of 
ESO~137-001, in this section we will summarize and compare the 
fractions of different gas phases in the tail as a whole, as well as 
search for trends along the tail by studying local variations in the 
observed regions.

\subsection{The whole tail of ESO~137-001}

\begin{table}
\centering
\caption[]
{Masses of individual ISM components in ESO~137-001.}
\label{masses}
\begin{tabular}{lccccc}
\hline
\hline
\noalign{\smallskip}
Source & cold H$_2$ & warm H$_2$ & \ion{H}{1} & H$\alpha$ & X-rays\\
       & ($10^8 M_\odot$) & ($10^8 M_\odot$) & ($10^8 M_\odot$) &
       ($10^8 M_\odot$) & ($10^8 M_\odot$)\\
\noalign{\smallskip}
\hline
\noalign{\smallskip}
body & $<11$\tablenotemark{a} & - & $<5$\tablenotemark{b} & 
-\tablenotemark{c} & 0.8\\
tail & $>11$\tablenotemark{a} & $>0.4$ & $< 24$ & 1.9  & $11$ \\
\noalign{\smallskip}
001-A & 6.4 & $\sim 0.2$ & $<3.5$ & $0.8$ & 1.3 \\
001-B & 3.4 & $\sim 0.1$ & $<3.5$ & $0.5$ & 0.9 \\
001-C & 1.8 & -          & $<3.5$ & $0.7$ & 1.4 \\
\noalign{\smallskip}
\hline
\noalign{\smallskip}
\end{tabular}
\tablecomments{Masses of individual ISM components in ESO~137-001 and its stripped
tail. The volume filling factor of 0.05 and 1 are assumed for the
H$\alpha$ and X-ray emitting gas, respectively (see the text). }
\tablenotetext{1}{It is probable that some fraction of CO emission
in the main body pointing is coming from the encompassed off-disk inner
tail \ion{H}{2} regions.}
\tablenotetext{2}{The ATCA upper limit of $5\times 10^8~M_\odot$ per
$30''$ beam would correspond to \ion{H}{1} mass upper limit in the
optical area of the disk of $\sim 1\times 10^9~M_\odot$. However, we
stick to the lower value as \ion{H}{1} from outer disk radii has likely been
stripped.}
\tablenotetext{3}{The H$\alpha$ disk is much much brighter than the diffuse 
H$\alpha$ emission and is considered to be mainly mixture of many 
\ion{H}{2} regions.}
\end{table}

In addition to our APEX detection of more than $1\times 10^9~M_\odot$
of cold molecular gas in the three observed pointings in the tail of
ESO~137-001, current amounts of other gas phases in the tail have been
known at least to upper limits from previous observations:
\begin{itemize}
\item[--] About $1\times 10^9~{\rm f_{X}^{1/2} }~M_\odot$ of soft 
($0.5- 
2$~keV) X-ray emitting gas was revealed by \textit{Chandra} and
XMM-\textit{Newton} imaging in the two tails \citep{sun2006, sun2010},
where ${\rm f_X}$ is the volume filling factor of the X-ray emitting
gas that is expected to be close to 1 \citep{sun2010}.
\item[--] No \ion{H}{1} was detected to a limit of $5\times 
10^8~M_\odot$ per
$30''$ beam with ATCA \citep[Australia Telescope Compact Array,
][]{vollmer2001} which translates to an upper limit of $\sim 2\times 
10^9~M_\odot$ for the whole tails, assuming they cover about 4~times
the ATCA beam size.
\item[--] Up to $5\times 10^8~{\rm f_{H\alpha}^{1/2} }~M_\odot$ of 
ionized, H$\alpha$ emitting diffuse gas was found in the tails 
\citep{sun2007}, assuming the H$\alpha$ tail is approximated by a 
cylinder 40~kpc long with a diameter of 3.5~kpc, and where ${\rm 
f_{H\alpha}}$ is the volume filling factor of the H$\alpha$ gas that is 
expected to be low, as suggested from numerical simulations of ram 
pressure stripping where bright H$\alpha$ emission is produced at the 
edges of dense neutral clouds \citep{tonnesen2011}. 
For ${\rm f_{H\alpha}}\approx 0.05$ \citep[see e.g., Sec.~8.4 
in][]{jachym2013}, the upper limit on the amount of H$\alpha$ emitting 
gas is about $1\times 10^8~M_\odot$.
\item[--] \textit{Spitzer} observations revealed about $4\times 
10^7~M_\odot$
of warm ($130-160$~K) H$_2$ in the galaxy and the inner 20~kpc of the
tail \citep{sivanandam2010}.
\end{itemize}

The total gas mass in the tail thus is $2\times 10^9\lesssim {\rm 
M_{gas}}\lesssim 4\times 10^9~M_\odot$, including the ATCA \ion{H}{1}
upper limits, while the total gas mass in the disk of ESO~137-001 is
$\sim 1\times 10^9~M_\odot$ (see Table~\ref{masses}). It means that the
amounts of ram pressure stripped gas possibly exceed the current gas 
mass left in the disk. Since the original (pre-stripping) gas content 
of the galaxy estimated from its stellar mass and an assumption of a 
typical gas to stellar mass ratio is $\sim (0.5-1)\times 
10^{10}~M_\odot$ \citep{sun2010, sivanandam2010}, it also means that
the observed amount of stripped gas nearly accounts for the missing gas
from the disk. The mass ratio of the most abundant phases in the whole 
tail, the cold-to-hot, (H$_2$+\ion{H}{1})-to-X-ray, gas spans from 
about $1:1$ to $\lesssim 3:1$, assuming the Galactic CO-to-H$_2$ 
conversion factor and taking into account the \ion{H}{1} upper limit. 
The amounts of cold and hot gas in the tail are thus large and similar.

Our observations show for the first time that H$_2$, H$\alpha$, and 
X-ray emission can be at observable levels in a single ram pressure
stripped tail at a large scale of several tens of kpc. This supports
predictions of numerical simulations \citep{tonnesen2011,
tonnesen2012} that further suggested that the amounts of
X-ray emitting gas and the star-forming cold gas in the tail depend
strongly on the surrounding ICM pressure. 
In their model with the thermal ICM pressure close to
that at the location of ESO~137-001 (but with a lower ram pressure),
the ratio of cold-to-hot components in the tail is about $1:1.4$, and 
it drops to $\sim 1:9$ for about a two-times higher thermal pressure.
Note however, that in \citet{tonnesen2011} the cold component 
corresponds to gas in the temperature range of $3\times 10^2- 10^4$~K, 
and the hot component to $7\times 10^5- 4\times 10^7$~K, i.e., $0.06- 
3.45$~keV, thus not completely consistent with the observations. 
They also estimated that in the Norma cluster, the radius at which the
ICM pressure falls below the minimum value needed for a tail to have
observable X-ray emission, is $\sim 250$~kpc. This is very
close to the projected distance of ESO~137-001 from the Norma center,
which suggests that the galaxy may have lit up in soft X-rays rather
recently.

\subsection{Systematic gradient along the tail}\label{gradient}
\begin{figure}
\centering
\includegraphics[height=0.47\textwidth,angle=270]{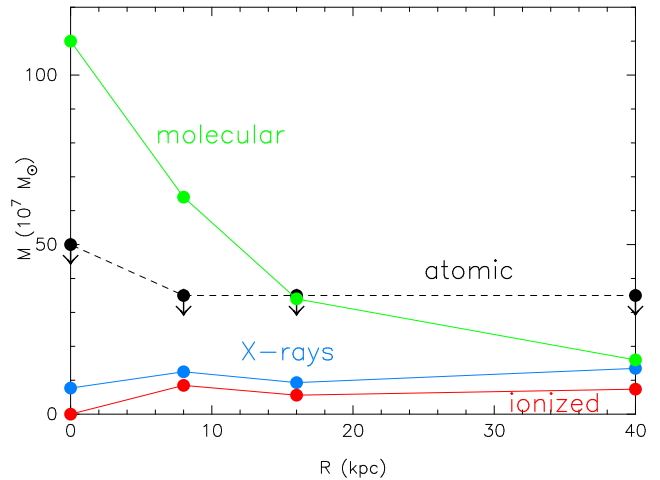}
\caption{
Masses of individual gas phases in the four observed regions (beams of
$\sim 8$~kpc FWHM) as a function of projected distance from the
galaxy -- molecular gas ({\it green}), diffuse H$\alpha$ ({\it red}), 
X-rays
({\it blue}), and upper limits on \ion{H}{1} ({\it black}). A low
H$\alpha$ filling factor of 0.05 is used (see the text). Molecular gas
mass depends on the correction factor to the Galactic value of the
CO-to-H$_2$ conversion factor.
}\label{FigGasComp}
\end{figure}

In order to understand the evolution of gas in the tail, it is of great
interest to know how the balance between the different gas phases
changes along the tail. It is expected that the ram pressure stripped
cool gas either heats up by the surrounding hot ICM or cools via
radiative cooling and transforms into molecular gas. In
Table~\ref{masses} and in Fig.~\ref{FigGasComp}, we summarize the
amounts of different gas phases in the observed tail regions (APEX
beams) as a function of the projected distance from the disk. We note
that the lines in Fig.~\ref{FigGasComp} connecting the four data points
are intended to guide the eye and do not reflect real mass profiles
between the observed positions. Fig.~\ref{FigGasComp} clearly shows
that the farther from the disk, the amount of molecular gas decreases
while the amounts of diffuse ionized and X-ray emitting components stay
roughly constant.
We summarize the local mass ratios of individual gas phases along the
tail:
\begin{itemize}
\item[--] The ratio of cold-to-hot (H$_2$-to-X-ray) masses decreases 
from
about 5 in the 001-A region to $\sim 1$ in the 001-C region, assuming
a constant CO-to-H$_2$ conversion factor.
\item[--] Also the ratio of cool-to-warm ionized (H$_2$-to-H$\alpha$) 
components along the tail decreases from $\sim 8$ in the 001-A region 
to $\sim 3$ in 001-C, assuming that the H$\alpha$ volume filling factor 
does not change.
\item[--] The ratio of hot-to-warm ionized (X-ray-to-H$\alpha$) masses 
is quite similar in all three regions and close to $\sim 2$.
\item[--] The warm-to-cold molecular gas ratio can be also followed. 
The region defined in \citet{sivanandam2010} as ``tail'' covers about a 
half of the 001-A pointing and a quarter of the 001-B pointing. It 
contains $3.6\times 10^7~M_\odot$ of warm H$_2$, while $< 4\times 
10^8~M_\odot$ of cold H$_2$. Thus, the warm to cold (mid-IR-to-CO) 
H$_2$ mass fraction is $\gtrsim 0.1$.
\end{itemize}

The ratio of warm-to-cold molecular gas $>0.1$ in the inner tail of
ESO~137-001 is higher than in star-forming galaxies, where it usually
is $0.004-0.04$ \citep[][]{guillard2012, nesvadba2010, roussel2007}.
The observed value is close to ratios measured in cool cores of cluster
and group central galaxies or in environments where molecular gas is
shock or cosmic rays heated rather than UV-heated. For example, in 
intra-galactic medium of Stephan's Quintet the ratio of warm-to-cold 
H$_2$ was measured to $\approx 0.3$ by \citet{guillard2012}, who also 
suggested that a high heating rate in the H$_2$ gas is probably 
maintained by turbulence invoked by galaxy-tidal arm collision. The 
observed value is consistent with shock heating \citep[as suggested 
by][]{sivanandam2010} and possibly heat conduction from ICM as other 
sources of heating that must exist in the tail, other than star 
formation. It is likely that the diffuse component of H$\alpha$ in the tail is not
associated with star formation but is powered by shocks or other 
processes, such as heat conduction \citep{sun2007}.
For example, shocks from ram pressure were proposed to be 
responsible for elevated levels of radio continuum emission, as well as 
for elevated ratios of warm H$_2$/PAH in Virgo spirals being stripped
\citep[][respectively]{murphy2009, wong2014}. Consequently, the CO 
emission in the tail of ESO~137-001 may trace not only cold molecular 
gas but also warmer gas \citep{guillard2012} and thus the actual amount 
of cold H$_2$ in the gas-stripped tail could be lower than 
corresponding to the standard CO-to-H$_2$ conversion factor \citep[e.g.,][]{bolatto2013}.

\begin{table}
\centering
\caption{
The H$\alpha$ and X-ray fluxes and masses in the observed regions.
}
\label{Ha_X_table2}
\begin{tabular}{lcccccc}
\hline
\hline
\noalign{\smallskip}
Source & $F_{0.5-2~{\rm keV}}$ & $M_{\rm X}$ & $F_{\rm H\alpha}$ &
$F_{\rm H\alpha}'$ & $M_{\rm H\alpha}$ & $\frac{F_{\rm X}}{F_{\rm 
H\alpha}'}$\\
\noalign{\smallskip}
\hline
\noalign{\smallskip}
ESO~137-001 & 2.7 &  7.7 & 11.7 & -   & -  & -   \\
001-A & 2.3 & 12.5 &  3.0 & 1.9 & 38 & 1.3 \\
001-B & 1.3 &  9.3 &  1.1 & 0.9 & 25 & 1.5 \\
001-C & 2.1 & 13.5 &  1.1 & 1.1 & 33 & 1.9 \\
\noalign{\smallskip}
\hline
\noalign{\smallskip}
\end{tabular}
\tablecomments{The H$\alpha$ and X-ray fluxes and corresponding masses
in individual observed points. Fluxes of the diffuse H$\alpha$-emitting
gas including the \ion{H}{2} regions ($F_{\rm H\alpha}$) and without the
\ion{H}{2} regions ($F_{\rm H\alpha}'$) are given. About 40\% of the 
H$\alpha$ diffuse flux in 001-A is from the galaxy and 001-B. The fluxes 
and masses are given in $10^{-14}$~erg\,s$^{-1}$\,cm$^{-2}$, and $10^7 
f^{1/2}~M_\odot$, respectively, where $f_{\rm H\alpha}$ and $f_{\rm X}$ 
stand for the respective filling factor. The tail was approximated by a 
cylinder with a diameter of 4.8~kpc (for 001-A and 001-B) and 5.5~kpc 
(for 001-C).
}
\end{table}

The H$\alpha$ and X-ray fluxes and masses derived for the
APEX apertures are listed in Table~\ref{Ha_X_table2}. The H$\alpha$
fluxes were derived using the
case B recombination theory and we didn't assume any intrinsic
extinction for the diffuse H$\alpha$ emission (although there may be as
there is dust in the tail). A large uncertainty comes from the
H$\alpha$ emissivity and the filling factor. The X-ray masses are
uncertain from the spectral model and the filling factor.

The positional coincidence of the H$\alpha$, X-ray, and CO emission in
the tail of ESO~137-001 suggests that different gas components mutually
interplay and possibly exchange energy. The origin of the observed soft
X-ray emission in the tail is presumed to arise at the mixing
interfaces of cold stripped ISM and the surrounding hot ICM
\citep{sun2010, tonnesen2011}.
The drop of molecular gas amounts along the tail could point to
decreasing mean gas density in the tail with increasing distance
(for example due to more acceleration at larger distances that makes
the gas more spread out in the tail direction in its outer parts), to
long timescale of heating, or to a recent increase in the ram pressure
stripping strength and stripping rate that has pushed more dense gas
to the inner tail regions only (see Section~\ref{rps}). We note that
our CO observations may be biased by targeting only the H$\alpha$ 
peaks.
Also, changing X-factor along the tail could to some extent contribute
to the decrease of molecular gas. Even so, in ESO~137-001, it is the
first time that we can follow in rough outline the distribution of cold
molecular gas in a ram pressure stripped tail.

\section{Ram pressure stripping of ESO~137-001}\label{rps}
The previous thorough analysis of \citet{sivanandam2010, sun2007,
sun2010} showed that ESO~137-001 is a clear case of ram pressure
stripping, which is consistent with its high \ion{H}{1}-deficiency
$>10$, the undisturbed morphology of its stellar disk, and the presence
of a prominent one-sided tail.
Other indicators suggest that the effects of ram pressure on the galaxy
may have been strong: ESO~137-001 is projected very close to the center
of the Norma cluster that moreover is a rich and massive cluster. Thus,
local ICM density is presumably high, as well as the galaxy's orbital
velocity.
In addition, the rather low mass of ESO~137-001 makes it easier to
strip. Its rotation velocity of $\sim 110$~km\,s$^{-1}$ is lower than
rotation velocities of the most studied Virgo ram pressure stripped
galaxies with typically $\upsilon_{\rm rot}\sim 130$~km\,s$^{-1}$
\citep[NGC~4522, NGC~4402;][]{kenney1999, crowl2005} up to $\sim 
170$~km\,s$^{-1}$ \citep[NGC~4330;][]{abramson2011}.

\subsection{Possible orbits}
In order to estimate the ram pressure stripping history of ESO~137-001,
we will first numerically study its possible orbits in the Norma
cluster. Such orbits need to be consistent with the galaxy's current
position on the sky, the line-of-sight velocity, and the plane of the 
sky velocity direction (indicated by the projected tail 
direction)\footnote{The position angle of the (main) tail varies along 
its length -- in the inner parts, both young stellar streams in the HST 
image and the X-ray emission show PA$\sim 316^\circ$, while in the 
outer tail it is $\sim 306^\circ$. We measure PA from North through 
West.}. The free parameters of the model are the current l-o-s distance
relative to the cluster center and velocity component in the plane of
the sky. According to the measured absence of a radial velocity
gradient along the tail, most of ESO~137-001's orbital motion must be
happening close to the plane of the sky, and the actual distance of the 
galaxy from the cluster center is thus possibly close to the projected 
one. The orientation of the gas-stripped tail pointing away from the 
cluster center then suggests that the galaxy is currently approaching
pericenter.

The gravitational potential of the Norma cluster can be modeled with a
Navarro-Frenk-White profile \citep[NFW,][]{navarro1996} with parameters
given in Table~\ref{TabModel}. This yields a cluster with $M_{\rm vir}= 
1.1\times 10^{15} M_\odot$ and $R_{\rm vir}= 2.2$~Mpc and the
distribution of mass with radius that is well consistent with the
values measured by \citet{woudt2008}. Since the Norma cluster is not 
relaxed, its dark matter distribution may be in reality bimodal, 
following the two interacting subclusters around giant ellipticals 
ESO~137-008 and ESO~137-006 (see Fig.~\ref{FigNorma}). ESO~137-001 
however lies outside of the central merging region. Also the 
distribution of ICM in the NW part of the cluster looks more symmetric 
than the opposite side where the X-ray emission is elongated. In our 
modeling ESO~137-006 was considered the cluster center (see 
Fig.~\ref{FigNorma}) since the peak of the galaxy density distribution 
is located only $\sim 3'$ from it \citep{woudt2008}.

\begin{table}[t]
\centering
\caption[]
{Parameters of the cluster and galaxy models. }
\label{TabModel}
\begin{tabular}{lll}
\hline
\hline
\noalign{\smallskip}
cluster: & $r_\mathrm{s}=346$~kpc & $c= 6.2$\\
         & $\rho_{\rm ICM, 0}= 4.82\times 10^{-24}$~kg\,m$^{-3}$ & $r_c=184.5$~kpc\\
         & $\beta=0.555$ &\\
\noalign{\smallskip}
\hline
\noalign{\smallskip}
galaxy:  & $M_\mathrm{d}= 1\times 10^{10}~M_\odot$ & $a_\mathrm{d}= 
3$~kpc\tablenotemark{a} \\
         & $M_\mathrm{b}= 1\times 10^9~M_\odot$ & $a_\mathrm{b}= 0.5$~kpc  \\
         & $M_\mathrm{h}= 7\times 10^{10}~M_\odot$ & $a_\mathrm{h}= 10$~kpc  \\
\noalign{\smallskip}
\hline
\noalign{\smallskip}
\end{tabular}
\tablecomments{Parameters of the cluster and galaxy models used for 
semi-analytic calculations of ram pressure stripping. Cluster 
parameters are from \citet{sivanandam2010}.}
\tablenotetext{1}{measured by \citet{sun2007}}
\end{table}

We cover a wide range of values of the free parameters: the velocity 
component $\upsilon_{xy}$ in the plane of the sky spans from 
1000~km\,s$^{-1}$ to 4000~km\,s$^{-1}$ and the l-o-s distance $z$ 
relative to the cluster center from $-600$~kpc to 600~kpc (although 
negative and positive distances are almost equivalent due to only small 
radial velocity component). For a given 
velocity the trend is the following: with increasing l-o-s distance $z$ the 
pericenter distance grows, and the apo-to-pericenter ratio decreases. 
Cosmological simulations \citep[e.g.,][]{wetzel2011} have suggested
that galaxies at $z=0$ infalling to massive ($M=10^{14}-10^{15} 
M_\odot$) host halos occur preferentially on radial orbits with an
average eccentricity of $e= 0.85$, and thus a mean apo-to-pericentric
ratio $r_{\rm apo}/ r_{\rm peri}= (1+e)/ (1-e)$ of $\sim 12$. Mean
pericenter and apocenter distances are $r_{\rm peri}= 0.13~R_{\rm vir}$
and $r_{\rm apo}= 1.6~R_{\rm vir}$, respectively. Typical turn-around
radii (where the characteristic infall velocity is equal to the Hubble
flow) of radially infalling galaxies in clusters are thus several~Mpc
\citep[e.g.,][]{zu2013}. ESO~137-001 is believed to be on its first
infall.

\begin{figure}
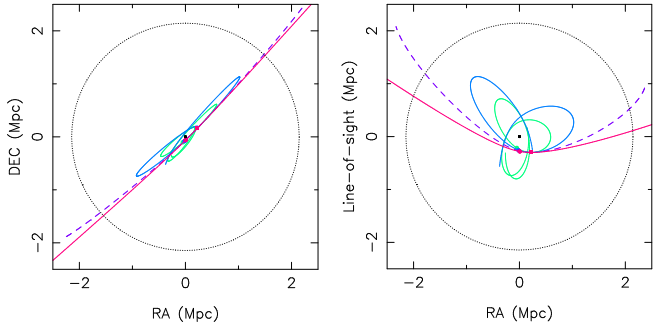

\centering
\includegraphics[height=0.23\textwidth,angle=270]{./f6a.ps}
\hspace{0.1cm}
\includegraphics[height=0.23\textwidth,angle=270]{./f6b.ps}
\caption{
RA--Dec and RA--line-of-sight views of the model orbits with the plane 
of the sky velocity component changing from 2100~km\,s$^{-1}$ to 
3900~km\,s$^{-1}$, with a 600~km\,s$^{-1}$ step, and the l-o-s distance 
of the galaxy of $300$~kpc relative to the cluster center. The cluster 
virial radius of $\sim 2.2$~Mpc is shown with the dashed circle. 
The sections of orbits shown correspond to a travel time of $\sim 5.5$~Gyr. 
The square marks the current position of the galaxy and the dots the 
pericenters. The 3300~km\,s$^{-1}$ (dashed) orbit is our fiducial orbit 
used for ram pressure stripping modeling. 
}\label{FigOrbits}
\end{figure}

Although the velocity dispersion of $\sim 949$~km\,s$^{-1}$ measured
within $\sim 0.67$~Mpc radius of the Norma cluster \citep{woudt2008} 
suggests a 3D orbital velocity of $\sqrt{3}\sigma= 1644$~km\,s$^{-1}$, 
all our model runs for ESO~137-001 with velocities lower than $\sim 
3000$~km\,s$^{-1}$ are too compact and do not cross the cluster virial 
radius. 
This is illustrated in Fig.~\ref{FigOrbits} where model orbits with the 
current plane of the sky velocity component in the range from 
2100~km\,s$^{-1}$ to 3900~km\,s$^{-1}$ are shown. 
Thus to reach the above typical
orbital parameters of infalling galaxies in cosmological simulations,
our modeling suggests ESO~137-001 to have a rather high orbital
velocity $\gtrsim 3000$~km\,s$^{-1}$. This is also supported by
simulations of \citet{tonnesen2008} that suggested that the relative
galaxy-ICM velocities within inner $\sim 300$~kpc of a large $M_{200}= 
6\times 10^{14}~M_\odot$ cluster are larger than $\sim 
2500$~km\,s$^{-1}$. 

As a fiducial orbit for further ram pressure modeling we select the one 
with $\upsilon= 3300$~km\,s$^{-1}$ and the line-of-sight distance 
relative to the cluster center of $z=300$~kpc (see the purple curve in 
Fig.~\ref{FigOrbits}) for which the pericenter distance and the 
peri-to-apocenter ratio are consistent with the mean values predicted 
for infalling galaxies from cosmological simulations. Along the 
fiducial orbit the galaxy currently is about 100~Myr before pericenter.

\subsection{Effects of ram pressure}
We will follow the effects of ram pressure on ESO~137-001 along the
fiducial model orbit. By means of semi-analytic calculations we
estimate how ISM parcels with different column densities react to a
time varying ram pressure. We thus neglect all hydrodynamic effects,
such as Kelvin-Helmholtz (KH) and Rayleigh-Taylor (RT) instabilities or
gas compression that can accompany the dynamical ram pressure. The 
acceleration due to local ram pressure may be expressed as
\begin{equation}\label{EqRP}
\frac{d\upsilon}{dt}\, \Sigma_{\rm ISM}= \rho_{\rm ICM}\, (\upsilon- 
\upsilon_0)^2 - \frac{\partial \Phi}{\partial z}\, \Sigma_{\rm ISM}, 
\end{equation}
where $\Sigma_{\rm ISM}$ is the mean column density of an ISM parcel, 
$(\upsilon-\upsilon_0)$ is the vertical component of its relative 
velocity with respect to the surrounding ICM, and $(\partial 
\Phi / \partial z)\, \Sigma_{\rm ISM}$ is the gravitational restoring 
force of the disk + bulge + halo galaxy components in the vertical 
direction. We then numerically solve the equation of motion 
(Eq.~\ref{EqRP}) along the fiducial orbit for different values of 
$\Sigma_{\rm ISM}$ and for different disk radii. We have not accounted 
for the changing wind angle along the orbit and instead use a face-on
wind\footnote{With the small relative radial velocity of the galaxy and
its high inclination, ram pressure is expected to be operating on the
galaxy close to face-on ($\lesssim 30^\circ$).}.

To model the ICM distribution we follow
\citet{bohringer1996} who fitted the western less disturbed part of the
A3627 cluster with a $\beta$-model centered on the X-ray peak located
about $5.5'$ west from ESO~137-006. 
The $\beta$-profile parameters are given in Table~\ref{TabModel},
together with the parameters of the model galaxy. For ESO~137-001, the
$K$-band velocity-luminosity relation $M_K=-22.5$~mag (NED) yields a
rotation velocity of $\sim 110-120$~km\,s$^{-1}$ \citep{courteau2007}.
Assuming a flat rotation curve this corresponds to a dynamical mass of
$\sim 3.5\times 10^{10}~M_\odot$ within the observed stellar disk
radius of $\sim 40''\cong 13$~kpc \citep{sivanandam2010}. We assume a
simple Plummer dark matter halo with the total mass of $\sim 7\times 
10^{10}~M_\odot$ and the scale radius of 10~kpc, although the halo 
parameters are somewhat uncertain and the simulation is sensitive to 
them.

Along our fiducial orbit the galaxy is currently about 100~Myr before 
pericenter but due to the offset of ICM distribution from ESO~137-006, 
it is only about 60~Myr before peak ram pressure. Correspondingly, the 
current ram pressure has a high value of
$\sim 1.5\times 10^{-10}$~dyne\,cm$^{-2}$ and the peak ram pressure 
will reach $\sim 2.1\times 10^{-10}$~dyne\,cm$^{-2}$. The FWHM of the 
ram pressure time profile is $\sim 200$~Myr. 

\begin{figure}
\centering
\includegraphics[height=0.47\textwidth,angle=270]{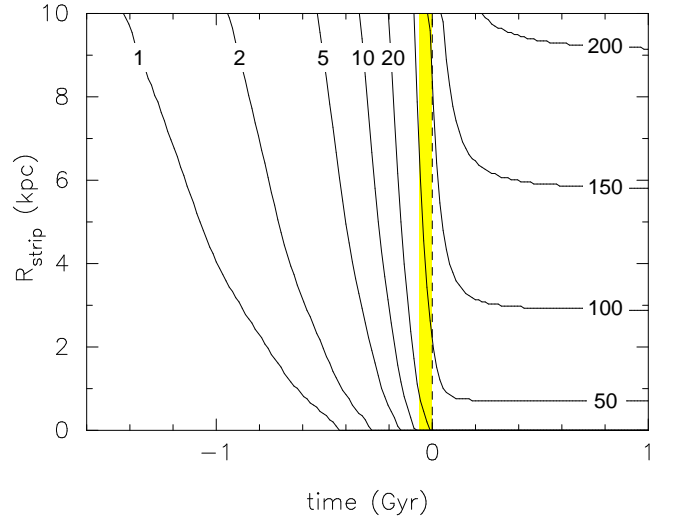}
\caption{
Effects of ram pressure along the fiducial orbit -- evolution of 
stripping radii of ISM components with different column densities 
(curves from left to right; values are in $M_\odot$\,pc$^{-2}$ units). 
The peak of ram pressure occurs at time 0~Gyr (dashed vertical line). 
Time remaining to the peak is indicated with the shaded area. 
}\label{FigRstrip-time}
\end{figure}

Figure~\ref{FigRstrip-time} displays the evolution of stripping radii 
for ISM components with different column densities along the fiducial 
orbit. We consider an ISM element to be stripped when its vertical 
velocity exceeds the escape speed from the galaxy. 
Fig.~\ref{FigRstrip-time} suggests that gas with $\Sigma_{\rm 
ISM}\lesssim 10~M_\odot$\,pc$^{-2}$ is currently completely stripped 
from the galaxy ($R_{\rm strip}= 0$~kpc), and that stripping has 
proceeded to denser gas (about $20-50~M_\odot$\,pc$^{-2}$) at larger 
disk radii. The figure also indicates the timescale of stripping.
As expected, lower-density gas begins to be accelerated earlier along 
the orbit than more dense clumps, and also stripping proceeds from 
outer radii of the disk inwards. It means that a mixture of different 
density gas forms in the tail since denser clumps from outer radii may 
be stripped to similar distances at the same time as less dense gas 
from inner radii. The timescale for stripping depends on the ISM 
parcel's column density and on its location in the disk. 
In Fig.~\ref{FigRstrip-time} one can also infer how dense ISM will be 
stripped from the galaxy in the future after it passes through the peak 
ram pressure. 

\begin{figure}
\centering
\includegraphics[height=0.47\textwidth,angle=270]{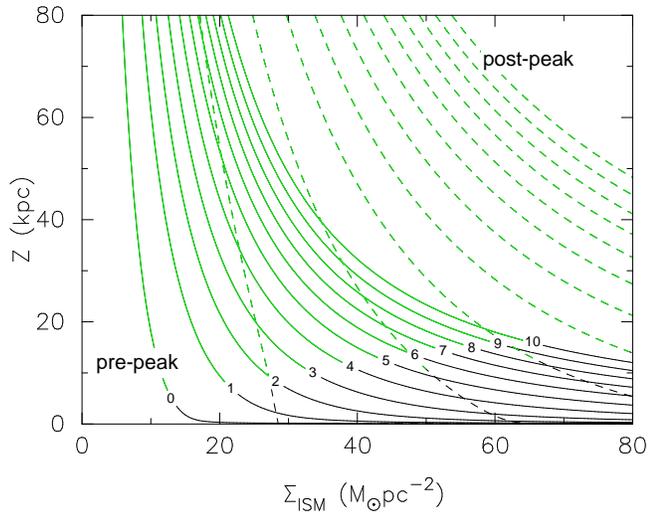}
\caption{
Effects of ram pressure along the fiducial orbit -- vertical distances 
to which can get ISM elements with different column densities from 
different disk radii ($0-10$~kpc; curves from left to right; see labels 
in kpc units) at current, $\sim 60$~Myr pre-peak time (solid lines), 
and for comparison symmetrically at $\sim 60$~Myr post-peak time 
(dashed lines). 
The radius labels separate the segments of the curves where ISM is 
stripped (green) or only pushed but still bound to the galaxy. 
}\label{FigSigISM-Z}
\end{figure}

To compare the predictions of our model with the observed length of the 
tail of ESO~137-001, Fig.~\ref{FigSigISM-Z} displays the vertical 
distances that ISM elements have reached as a function of their column 
density and radius in the disk (solid curves). It indicates that gas 
with the maximum density of $\sim 20~M_\odot$\,pc$^{-2}$ (depending on 
radius) could currently be stripped to large distances exceeding the 
observed length of the ESO~137-001's X-ray tail, while denser gas of up 
to $\sim 80-100~M_\odot$\,pc$^{-2}$ is only pushed to inner tail regions. 
More specifically, Table~\ref{TabStripping} gives column densities of 
elements stripped from different disk radii to vertical distances 
matching those of the whole tail length, and of the projected locations 
of the 001-C and 001-A regions. For comparison, situation in a 
symmetrical, 60~Myr post-peak time, i.e., about 120~Myr from now, is 
also shown in Fig.~\ref{FigSigISM-Z} (dashed curves). 

\begin{table}[t]
\centering
\caption[]
{Column densities of stripped ISM in our model.}
\label{TabStripping}
\begin{tabular}{l|ccc}
\hline
\hline
\noalign{\smallskip}
    & \multicolumn{3}{c}{Radius in the disk:}\\
Vertical distance & 2~kpc & 6~kpc & 10~kpc\\
\noalign{\smallskip}
\hline
\noalign{\smallskip}
tail (80~kpc) & 10 & 17 & 20\\
001-C (40~kpc) & 15 & 25 & 35\\
001-A (10~kpc) & 25 & 60 & 90\\
\noalign{\smallskip}
\hline
\noalign{\smallskip}
\end{tabular}
\tablecomments{
Approximate column densities (in $M_\odot$\,pc$^{-2}$ units) of ISM 
stripped (or pushed) along our fiducial orbit from different disk radii 
to vertical distances distinctive for the observed tail of ESO~137-001. 
The results follow from Fig.~\ref{FigSigISM-Z}.
}
\end{table}

Our calculations show the effects of ram pressure on ISM with different 
column densities and suggest that rather dense clumps may be 
stripped/shifted from the disk of ESO~137-001. 
However, real GMCs have strong internal density gradients, and ram 
pressure is thus most likely able to ablate the outer portions of a 
cloud rather than remove it entirely (even though the cloud's average 
density may be low enough to be displaced according to the model). 
Current observations of Galactic GMCs have suggested that column 
densities of dense cores of GMCs reach values of several hundreds 
of $M_\odot$\,pc$^{-2}$ while the mean column density over the whole 
clouds is rather low \citep[$\sim 42~M_\odot$\,pc$^{-2}$,][]{heyer2009}, 
much lower than previously estimated by \citet{solomon1987}. 
Fig.~\ref{FigRstrip-time} shows that ram pressure has likely been 
increasing substantially during last couple of 100~Myr and thus it is 
possible that ablation of clouds may have progressed into the central parts 
of some of them and effectively stripped them. 

The exceptional features of ESO~137-001 (in comparison with most of 
other known RPS galaxies), such as the bright and long X-ray tail, 
unprecedented SF activity downstream from the galaxy, as well as the wealth 
of CO emission in the tail, support the idea of a recent strong ram 
pressure stripping. This is also suggested by the presence of dust 
filaments revealed in the HST and {\it Herschel} imaging since typical 
column densities of the dust clouds that produce significant optical 
extinction are at least $\sim 5~M_\odot$\,pc$^{-2}$ 
\citep{abramson2014}. Very recently, \citet{ebeling2014} reported a 
discovery of several strongly ram pressure stripped galaxies falling 
into massive $z>0.3$ clusters. Similarly to ESO~137-001, these galaxies 
are located close ($\sim 90- 360$~kpc in projection) to the center of 
their host cluster. It is thus possible that also in these galaxies ram 
pressure stripping may have currently proceeded to denser ISM since 
\ion{H}{1} removal is expected to take place at larger distances from 
the cluster center along the orbit of the infalling galaxies.

\section{Discussion}
In the following sections we will identify possible traces of 
turbulence in the tail of ESO~137-001 from kinematic structure of the 
detected CO lines, and we will discuss turbulence as a possible 
mechanism suppressing star formation in the stripped gas. Since 
hydrostatic pressure in galactic disks is suggested to be the key 
parameter that determines the transformation between atomic and 
molecular gas, we will compare the ICM pressure operating on 
ESO~137-001 with values typical in galaxies. Further, we will discuss 
possible origins of the detected molecular gas in the galaxy's tail, as 
well as think about the evolutionary state of the ''orphan'' \ion{H}{2} 
regions occurring in the broad parts of the tail. 

\subsection{Multiple line components and the tail kinematics}\label{multiple}
While the ESO~137-001 main body CO line is only slightly asymmetric,
the tail spectral lines all show substructure. Using the CLASS routine
FIT we decomposed the spectra in the 001-A and 001-B positions into
several Gaussian components (see Fig.~\ref{Fig001} and
Table~\ref{TabRes}) showing that the kinematics of the CO-emitting gas
is rather complex, especially in the inner parts of the tail. This may 
reflect individual (large) condensations of dense gas possibly 
consistent with the observed distinct \ion{H}{2} regions (as already 
suggested in Fig.~\ref{Fig001} by blue bars showing radial velocities 
of the \ion{H}{2} regions). The molecular gas masses of the individual 
resolved components span a range of $\sim 5\times 10^7 - 4\times 
10^8~M_\odot$. The FWHMs of their Gaussian fits are of the order of 
$\sim 25- 64$~km\,s$^{-1}$, i.e., $\sim 11- 27$~km\,s$^{-1}$ velocity 
dispersions (since the relationship between FWHM of a Gaussian fit and 
the velocity dispersion is ${\rm FWHM}\approx 2.3548\,\sigma$). These 
rather small values are consistent with the ongoing star formation in 
the spatially associated regions (star clusters) since the gas is more 
concentrated and possibly self-gravitating. Much better spatial and 
spectral resolution would be needed to further resolve smaller 
entities, such as individual molecular clouds whose velocity 
dispersions are smaller, typically $\sim 10$~km\,s$^{-1}$, and their 
masses are $10^5- 10^6~M_\odot$, and diameters $10- 100$~pc 
\citep[e.g.,][]{solomon1987}.

In the observed tail regions, the CO emission is more extended in 
velocity than in the main body (see Fig.~\ref{Fig001} and 
Table~\ref{TabRes}), especially in the 001-B pointing where individual 
line components are clearly distinct. While some of the velocity spread 
could be due to the rotational motion of the disk that will survive 
into the (inner) tail, we expect turbulent motions driven to the 
stripped ISM by the high-speed interaction between the galaxy and the 
surrounding ICM to be important. A turbulent velocity field was 
observed to develop in the stripped gas in numerical simulations due 
to the KH instability that leads to formation of eddy-like structures 
in the tail \citep[e.g.,][]{tonnesen2010, roediger2005, roediger2006}. 
The typical rms turbulent velocity averaged over a galaxy's tail is 
predicted to be of the order of $\sim 100$~km\,s$^{-1}$ 
\citep[e.g.,][]{subramanian2006}. For example, in the Virgo cluster 
galaxy NGC~4388, \citet{yoshida2004} revealed turbulent kinematic 
structure of the ram pressure stripped ionized gas with abrupt velocity 
changes of up to $\sim 300$~km\,s$^{-1}$. Dense gas that we detect in 
CO line emission possibly does not reflect the whole range of turbulent 
motions present in the tail of ESO~137-001. 

The present observations allow us to roughly estimate the turbulent
kinetic energy carried by the CO emitting gas in the tail of
ESO~137-001.
From $E_{\rm kin}= 3/2 M_{\rm H_2} \sigma_{\rm CO}^2$, where $M_{\rm 
H_2}$ is the molecular gas mass derived from the CO(2-1) intensity and
$\sigma_{\rm CO}$ is the 1-dimensional, l-o-s velocity dispersion
measured from the (multiple) Gaussian fits of the CO(2-1) lines (see
Table~\ref{TabRes}), the kinetic energies for the 001-A, 001-B, and
001-C tail positions are about $11.6\times 10^{54}$~erg, $5.5\times 
10^{54}$~erg, and $7.4\times 10^{54}$~erg, respectively. In total, the
observed CO emitting gas in the tail carries $\sim 2.5\times 
10^{55}$~erg. For comparison, the main body position has the kinetic
energy of about $3.8\times 10^{55}$~erg, most of its linewidth however
comes from (ordered) rotational motion, not turbulent motion. This is
also partly true for the tail positions
as suggested from radial velocities of the \ion{H}{2} regions
\citep{sun2010}. The total turbulent energy in the CO emitting gas in
the tail thus may be lower than the above estimated value by a factor
of few. For comparison, we can estimate the thermal energy of the hot
\citep[$\sim 0.8$~keV;][]{sun2010} X-ray emitting plasma encompassed in
the three observed regions in the tail of ESO~137-001 from $E_{\rm 
therm}= 3/2 n_e k T$. Assuming the beam area $\sim 80$~kpc$^2$, the
tail l-o-s width $\sim 7.8$~kpc, and the electron density $n_e\sim 
0.007$~cm$^{-3}$ \citep{sun2010}, the carried thermal energy is $\sim 
7\times 10^{56}$~erg. Despite the limitations of the estimate, it
suggests that the thermal energy of the hot gas is by a factor of $\sim 
10$ larger than the turbulent energy of the CO emitting gas.

\subsection{Molecular gas fraction vs. local pressure in the tail}\label{pressure}
\begin{figure}
\centering
\includegraphics[clip=true,width=0.47\textwidth]{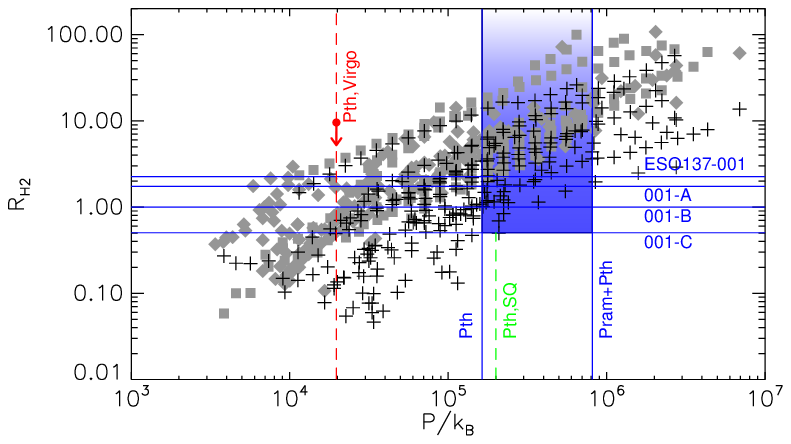}
\caption{
Molecular-to-atomic ratio, $R_{\rm H_2}= \Sigma_{\rm H_2}/ \Sigma_{\rm 
HI}$, as a function of midplane pressure (in K\,cm$^{-3}$) in
observations of nearby galaxies (squares -- \citealt{blitz2006};
diamonds -- \citealt{leroy2008}) and theoretical predictions of 
\citet[plus signs]{krumholz2009}. Lower limits on $R_{\rm H_2}$ in the 
observed regions of ESO~137-001 are shown, together with values of 
thermal and total (thermal + ram) ICM pressure on the galaxy. For 
comparison, values of thermal pressure of the ICM in the Virgo cluster 
(at the position of IC3418 galaxy; {\it red}), and of the hot 
intra-group plasma in the densest, X-ray brightest region in the 
Stephan's Quintet ({\it green}) are indicated \citep{jachym2013, 
guillard2009}. Image adapted from \citet{krumholz2009}.
}\label{FigR2PRESS}
\end{figure}

The molecular-to-atomic surface density ratio ($R_{\rm H_2}= 
\Sigma_{\rm H_2}/ \Sigma_{\rm HI}$) correlates in galaxies with the
interstellar pressure \citep{blitz2004, blitz2006}. In 
Fig.~\ref{FigR2PRESS} we compare lower limits on $R_{\rm H_2}$ measured 
in the observed regions of ESO~137-001 (corresponding to our APEX 
detections and the ATCA upper limits, see Table~\ref{masses}) with data 
from disks of nearby galaxies \citep{blitz2006, leroy2008} in a figure 
adopted from \citet{krumholz2009}. In the plot we further constrain the 
thermal and total (thermal + ram) ICM pressure operating on 
ESO~137-001, corresponding to X-ray $Chandra$ observations 
\citep{sun2010} and our simulations (Section~\ref{rps}). The 
surrounding ICM thermal pressure at the position of ESO~137-001 in 
A3627 is $\sim 1.8\times 10^{-11}$~dyne\,cm$^{-2}= 1.3\times 
10^5~k_{\rm B}$~K\,cm$^{-3}$. The resulting area in 
Fig.~\ref{FigR2PRESS} is filled with a color gradient indicating that 
if actual values of $\Sigma_{\rm HI}$ were known, the lower limits on 
$R_{\rm H_2}$ in ESO~137-001 would very likely shift to higher values 
and thus would cover values observed in disks of nearby galaxies.

Figure~\ref{FigR2PRESS} shows that the ICM (thermal and ram) pressure
at the location of ESO~137-001 in the Norma cluster is similar to 
pressures operating in the completely different environments of 
galactic disks. This offers an explanation for the detected bulk of 
molecular gas in the gas-stripped tail of the galaxy. For comparison, 
in Fig.~\ref{FigR2PRESS} we also indicate the value of the thermal 
pressure in the Virgo cluster, at the location of the ram pressure 
stripped dwarf galaxy IC3418. It is about an order of magnitude lower 
than the thermal pressure at ESO~137-001, thus suggesting much lower 
$R_{\rm H_2}$ ratios. Consistently, an upper limit on $R_{\rm H_2}$ in 
one of the observed regions in the outer tail of IC3418, where CO was 
only measured to upper limits and \ion{H}{1} was detected 
\citep{jachym2013, kenney2014} is shown with the red arrow. Further, in 
Fig.~\ref{FigR2PRESS} is also indicated the value of the thermal 
pressure of the hot plasma in the densest star-forming inter-galactic 
region in the X-ray brightest area (``ridge'') of the Stephan's Quintet 
\citep{guillard2009}. We note that the average thermal pressure in the 
Stephan's Quintet compact group is lower than the pressures in the 
Norma and Virgo clusters.

Although we find that the molecular-to-atomic gas ratios in the 
gas-stripped tail of ESO~137-001 may be consistent with values typical 
in galactic disks thanks to similar pressures in the two environments, 
the star formation efficiency in the tail is much lower than in the 
galaxies (as shown in Section~\ref{sfr}). Among the effects preventing 
star formation could be large scale turbulence driven from interaction 
with the surrounding ICM. Turbulence can provide further support 
against gravity in cold molecular clouds, leading to small SFEs.
\citet{sivanandam2010} detected pure rotational H$_2$ transitions S(0) 
through S(7) in the inner tail of ESO~137-001 suggesting that indeed 
during momentum transfer by ram pressure, shocks are driven into the 
ISM which increases its thermal and turbulent pressure. Moreover, they 
showed that H$_2$ S(1) emission is (in the inner tail) much more 
extended than $8~\mu$m and $24~\mu$m emission, indicating that star 
formation proceeds only in a part of the stripped molecular gas. This 
may suggest that most of the molecular gas detected in the stripping 
tail of ESO~137-001 is most probably not-self-gravitating due to 
turbulence and surviving thanks to a lack of photodissociating UV 
radiation. Also, magnetic fields could prevent the collapse of molecular 
clouds.

\subsection{Origin of H$_2$ in the tail}
In this section we qualitatively discuss the origin of the molecular
gas in the tail of ESO~137-001.
The total H$_2$ mass of more than $10^9~M_\odot$ revealed within the
three APEX apertures in the gas-stripped tail of the galaxy is a large
amount possibly exceeding the original H$_2$ mass in the galaxy. From
the scaling relation deduced for example from the COLDGASS survey
\citep{saintonge2011} or from typical molecular gas fraction of $\sim 
0.1$ in an Sc galaxy \citep{obreschkow2009}, an average H$_2$ mass of
$\sim (0.5- 1)\times 10^9~M_\odot$ is expected in an unperturbed galaxy
with the stellar mass of $\sim 10^{10}~M_\odot$, although such estimates
are only approximate (within a factor of few).

The origin of the molecular gas detected in the tail may be twofold.
Either it has entirely formed in situ out of stripped atomic gas, or
some pre-existing molecular gas contributed to it. For the in-situ
formation scenario, the density of the stripped gas plays a role since it
determines the timescale for condensation and H$_2$ formation
following an inverse relation \citep{guillard2009}. Higher-density 
clumps can then radiatively cool down more easily and eventually form
molecular gas, while the low-density stripped phase is compressed by the
ICM, starts to mix with it, and likely accounts for the X-ray emitting
hot gas in the tail \citep{tonnesen2011}. Our simplified calculations
in Section~\ref{rps} suggested that ram pressure operating on
ESO~137-001 has been strong enough to shift (or even strip) ISM clumps
with column densities exceeding values typical in some molecular
clouds. Since stripping proceeds (for a given disk radius) from low to
high density gas, distant tail regions are more likely to host H$_2$
that formed in situ from a lower density gas phase, while inner tail
may contain also directly stripped dense, possibly molecular ISM
clumps.

The question arises whether the molecular phase can survive the effects of
ram pressure and thus directly contribute to the observed CO surface
brightness. Molecular clouds are affected in several ways --
photodissociated by the UV radiation, X-ray dissociated 
\citep[as in X-ray dissociation regions, XDRs,][]{maloney1996}, 
disrupted and ablated by
hydrodynamic effects connected to the ram pressure stripping, or
dissolved due to their limited lifetime that typically is $\sim {\rm 
few}\times 10^7$~yr \citep[e.g.,][]{murray2011}. However, it can be
shown that the timescales of the hydrodynamic processes are long and
may be suppressed by magnetic fields \citep{yamagami2011}, as well as
the UV radiation that is the dominant removal process in the stellar
disks is presumably absent in the tail environment. Numerical
simulations also suggested that high ram pressures compress ISM clouds
making them more resistant to ablation \citep{tonnesen2009}.
Consequently, it is possible that molecular clouds could survive
through at least $\sim 10^8$~yr and indeed get by ram pressure to the
(inner) tail. \citet{sivanandam2010} also
suggested that the original molecular phase in the disk may be
dissociated by the shock induced by the supersonic ram pressure. Such a
shock-dissociated molecular gas however is expected to reform in the
tail if dust survives \citep[e.g.,][]{guillard2009}.

The presence of dust is crucial for H$_2$ formation. The HST (see
Fig.~\ref{FigPoints}, right panel; Sun et al., in prep.) and {\it 
Herschel} SPIRE (Sivanandam et al., in prep.) imaging indeed revealed
in ESO~137-001 a dust trail emanating from the galaxy that is coaligned
with its gas-stripped tail up to $\sim 40$~kpc.
In summary, we suggest that the two scenarios of H$_2$ origin likely
combine in the gas-stripped tail of ESO~137-001. The wealth of CO
emission observed in its inner tail may correspond to
directly stripped relatively dense ISM clumps that not only are able to
quickly form H$_2$ but also may possess molecular phase that survived
the effects of ram pressure. The CO emission detected in the outer tail
region is then more likely to originate from in situ formation from a
stripped diffuse atomic component.
Some \ion{H}{1} thus must be present in the tail of ESO~137-001. It is
possible that the ATCA observations \citep{vollmer2001} are at the
limit of detection. Data analysis is ongoing of deeper observations
with ATCA (Sun et al., in prep.).

\subsection{Orhpan \ion{H}{2} regions in the broad tail}
We have not observed the broad parts of the tail where ``orphan"
\ion{H}{2} regions not associated with any X-ray emission occur (see
the left panel of Fig.~\ref{FigPoints}). However, their H$\alpha$
luminosities are comparable to those in the central tail, suggesting
that a bulk of CO emission could be present in the broad tail too.
Calculations in Sec.~\ref{rps} suggest that these regions were
formed from relatively dense gas that was stripped from the outer disk
$\sim$ (few $\times$) $10^8$~yr earlier than equally dense gas from
the inner disk. This is due to the gravitational restoring force that
is weaker at larger galactocentric radii and thus a weaker ram pressure
is required to strip gas of a given surface density from the outer disk
than from the inner disk.
However, all \ion{H}{2} regions that
are spread through the tail are of a similar age ($< 10$~Myr). This
suggests that star-forming \ion{H}{2} regions are constantly forming
and dying away throughout the tail, possibly due to evolution of the
stripped gas. This is also seen in numerical simulations where the
exact temperature and density of the stripped gas determines how long
it takes before the gas cools and condenses into clouds, invoking a
large spread in the height of stars above the disc
\citep{tonnesen2012}. It is not
clear why X-ray emission is absent in the broad tail. It may be due to
too much mixing that already lowered the density of the hot gas to the 
point where the X-ray emissivity is low \citep{tonnesen2011}.
The orphan \ion{H}{2} regions thus possibly represent a later 
evolutionary
stage in a gas-stripped tail after the gas supply from part of the disk
is exhausted. Then the once X-ray bright (least dense) gas is pushed to
larger tail distances while the dense, star-forming regions decouple 
from it. At earlier stages, when there is still gas from the disk supplying
the tail, the tail has a range of densities, and fast-moving low
density gas can be spatially coincident with dense lumps. This is
probably seen in the narrow tail.

\section{Conclusions}
We present new APEX $^{12}$CO(2-1) observations of ESO~137-001, a Norma
cluster spiral galaxy that is currently being violently transformed
from a late to an early-type by strong ram pressure stripping. A
prominent 80~kpc X-ray double-structure tail extends from the galaxy,
together with a 40~kpc H$\alpha$ tail. It also contains warm H$_2$
emission in the inner tail. ESO~137-001 is an excellent example of
extreme ram pressure stripping and may become completely gas-free in
the near future. Our APEX observations reveal large amounts of cold
molecular gas traced by $^{12}$CO(2-1) emission in the disk of
ESO~137-001 as well as in its gas-stripped tail. It is the first time
that a large amount of cold H$_2$ is found in a ram pressure stripped
tail. The main results of our analysis are:
\begin{enumerate}
\item More than $10^9~M_\odot$ of molecular gas was detected in the
three APEX 230~GHz apertures along the tail of ESO~137-001, including a
$\sim 40$~kpc distant intra-cluster region in its outer part where both
X-ray and H$\alpha$ emission peak. Although the X-factor is uncertain
in the special environment of a gas-stripped tail, we assume the
Galactic value because the metallicity measured in the tail is close to
solar ($\sim 0.6~Z_\odot$). As suggested by the high warm to cold 
(mid-IR to CO) molecular gas mass ratio measured in the inner part of 
the tail, the CO emission may trace not only cold but also warmer 
molecular gas. Consequently, the standard value of the X-factor 
could overestimate the H$_2$ content in the tail. Nevertheless, 
currently there may be more molecular gas in the tail of the galaxy 
than in its main body.
\item In the most distant ($\sim 40$~kpc) tail region, more than
$10^8~M_\odot$ of H$_2$ was revealed. Since the estimated tidal
truncation radius of the galaxy is 15~kpc, the detected molecular gas
may already be released from the gravitational well of the galaxy. The
detected amount is similar to typical molecular masses of tidal dwarf
galaxies. We speculate that a ram pressure dwarf galaxy (RPDG) could be
forming in this location. More detailed observations measuring 
the self-gravitation of the object are needed, though.
\item About $1\times 10^9~M_\odot$ of molecular gas was detected in the
APEX aperture centered on the main body of ESO~137-001. This is a
factor of at least $\sim 2$ less than the original molecular gas
content of the galaxy, as estimated from typical amounts expected in
unperturbed spiral galaxies of the same stellar mass.
\item Our observations show that H$_2$, H$\alpha$, and X-ray emission
can be at observable levels in a single ram pressure stripped tail. The
amounts of cold and hot gas in the tail are large and similar ($\sim 
10^9~M_\odot$) and together nearly account for the missing original gas
in the disk. Following our measurements, the amount of molecular gas
decreases along the tail, while masses of other gas phases (X-ray,
ionized) are roughly constant. This could correspond to a decreasing
mean gas density along the tail, to long timescales of heating, or to
increasing ram pressure that has been able to recently push denser gas 
to the inner tail regions.
\item The star formation efficiency was found to be very low in the
tail environment ($\tau_{\rm dep, H_2}> 10^{10}$~yr), while it is
consistent with other spiral galaxies in the main body of ESO~137-001.
This indicates that most of the stripped gas does not form stars but
remains gaseous and ultimately joins the ICM. Similarly low SFEs are
found for example in the outer disks of spiral galaxies where however
\ion{H}{1} is likely to represent most of the ISM and CO is mostly
undetected. This is in contrast to the CO bright tail of ESO~137-001 
where \ion{H}{1} was measured only to upper limits. Star clusters 
formed in the tail from the gas accelerated by ram pressure to high 
velocities exceeding the escape speed from the galaxy, will contribute 
to the intra-cluster light population.
\item The ICM thermal (+ ram) pressure at the location of ESO~137-001
in the Norma cluster is similar to midplane gas pressures that occur in
the (inner) disks of galaxies. Moreover, the lower limits on the
molecular-to-atomic gas ratio in the tail of ESO~137-001 (corresponding 
to our APEX detections and the ATCA \ion{H}{1} upper limits) are 
consistent with values measured in galactic disks. Nevertheless, the 
star formation efficiency in the tail is much lower than in the 
galaxies. We suggest that this is due to a low average gas density in 
the tail, or turbulence driven from interaction with the surrounding 
ICM, preventing the stripped cold gas from star formation. The elevated 
ratio of warm-to-cold molecular gas $>0.1$ in the inner tail of 
ESO~137-001 is close to ratios measured in cool cores of cluster and 
group central galaxies or in environments where molecular gas is heated 
by shocks or cosmic rays rather than UV-heated.
\item Our semi-analytic modeling of possible orbits of ESO~137-001 in
the Norma cluster indicate that the galaxy may be at a high velocity of
$\sim 3000$~km\,s$^{-1}$ in order to be consistent with a first infall
scenario. Consequently, the ram pressure experienced by ESO~137-001 
possibly has been strong enough to shift or strip from the disk 
some gas clumps with column densities exceeding values found in Milky 
Way typical molecular clouds. Such
dense gas in the tail can transform more readily into molecular gas
than stripped diffuse gas. Moreover, some fraction of the stripped gas
can survive in the molecular phase and contribute to the unprecedented
CO brightness of the gas stripped tail of the galaxy, especially in its
inner parts. The molecular gas detected in the outer tail is more likely
to originate from in situ transformation of stripped diffuse atomic gas.
\end{enumerate}
Future ALMA observations with high spatial resolution will enable us to
study ESO~137-001 in a great detail and to better understand the
process of cold gas stripping and mixing with the surrounding ICM, as
well as the star formation in the special environment of a ram pressure
stripped tail.

\section*{Acknowledgments}
We acknowledge support by the Czech Science Foundation grant
P209/12/1795, the project M100031203 of the Academy of Sciences of the
Czech Republic, and the institutional research project RVO:67985815.
This research was further supported under Australian Research Council's
Discovery Projects funding scheme (project number 130100664).
We thank our anonymous referee for helpful comments that improved the 
quality of this paper. 
This research has made use of the NASA/IPAC Extragalactic Database
(NED) which is operated by the Jet Propulsion Laboratory, California
Institute of Technology, under contract with the National Aeronautics
and Space Administration. We further acknowledge the usage of the
HyperLeda database (\url{http://leda.univ-lyon1.fr}).\\

{\it Facilities:} \facility{APEX (SHFI)}\\

\bibliographystyle{apj}
\bibliography{eso137}

\end{document}